\documentclass[11pt,titlepage,a4paper]{article}
\usepackage{bozhomac}	
\usepackage{bozhlogo}	
\usepackage{cite}	

\title{\bfseries	\vspace*{-1.678902345in}
{\huge On the action principle\\[1.6ex] in quantum field theory }
}

\vspace{1.7ex}

\author{
Bozhidar Z.\ Iliev
\thanks{Department of Mathematical Modeling,
Institute for Nuclear Research and \mbox{Nuclear} Energy,
Bulgarian Academy of Sciences,
Boul.\ Tzarigradsko chauss\'ee~72, 1784 Sofia, Bulgaria}
\thanks{E-mail address: bozho@inrne.bas.bg}
\thanks{URL: http://theo.inrne.bas.bg/$\sim$bozho/}
}

\date{
 \vspace{2.27ex}\ShortTitle{Action principle in QFT}	\\[0.27ex]
 \vspace{3.27ex}
\small
	\begin{tabular}{r@{$\colon\to~$}l}
 \vspace{0.09ex} Basic ideas	& July 21--22, 2001	\\[0.09ex]
 \vspace{0.09ex} Began		& August 8, 2001	\\[0.09ex]
 \vspace{0.09ex} Ended		& August 14, 2001	\\[0.09ex]
 \vspace{0.09ex} Initial typeset& August 15--August 20, 2001	\\[0.09ex]
 \vspace{0.09ex} Last update	& March 30, 2002	\\[0.09ex]
 \vspace{0.27ex} Produced	& \fbox{\today}	\\[0.27ex]
	\end{tabular} \\[1.27ex]
\normalsize
	\begin{tabular}{r@{$\colon~$}l}
\vspace{0.27ex} LANL xxx archive server E-print No. & hep-th/0204003\\[0.27ex]
	\end{tabular} \\[-0.27ex]
 \vspace{4.27ex}{\Huge\BOZHO}	\\[4.27ex]
 \vspace{0.27ex}\Subject{Quantum field theory}	\\[2.27ex]
	\begin{tabular}{r@{\hspace{0.512em}}|@{\hspace{0.512em}}l}
 \vspace{0.27ex}\MSC[2000]{81P99, 81Q99, 81T99}	
&
 \vspace{0.27ex}\PACS[2001]{02.90.+p, 03.70.+k, 11.10.Ef} 
	\end{tabular} \\[1.27ex]
 \vspace{0.27ex}\KeyWords{Quantum field theory, Action principle,
 Schwinger's action principle\\
 Action principles in quantum field theory, Conserved quantities\\
 Operators  of conserved quantities in quantum field theory\\
 Energy-momentum operator, Current operator\\
 Spin angular momentum density operator\\
 Euler-Lagrange equations, Field equations\\
 Derivative with respect to operator argument }	\\[0.27ex]
}

\newcommand{\opsi}{\overline{\psi}}	
\newcommand{\bpsi}{\Breve{\psi}}	
\newcommand{\tope}[1]{#1}	
\newcommand{\bv}{\Breve{v}}		

\newcommand{\Hil}{\mathcal{F}}		
\begin{document}		

\renewcommand{\thefootnote}{\fnsymbol{footnote}} 
\maketitle				
\renewcommand{\thefootnote}{\arabic{footnote}}   

\tableofcontents		


\begin{abstract}

	An analysis of the Schwinger's action principle in Lagrangian quantum 
field theory is presented. A solution of a problem contained in it is
proposed via a suitable definition of a derivative with respect to operator
variables. This results in a preservation of Euler\ndash Lagrange equations
and a change in the operator structure of conserved quantities. Besides, it
entails certain relation between the field operators and their variations
(which is identically valid for some fields, \eg for the free ones). The
general theory is illustrated on a number of particular examples.

\end{abstract}

\section {Introduction}
\label{Introduction}

	The paper deals with the following problems in Lagrangian quantum 
field theory: meaning of derivatives with respect to operator argument, order
of the operators in the structure of conserved quantities, and commutation of
the variations of the field operators in Schwinger's action principle. These
problems are reviewed, analyzed and their solution is proposed.

	The first two of the above problems are discussed in
Sect.~\ref{Sect2}. Sect.~\ref{Sect3} reviews the Schwinger's action principle
and some its consequences. Special attention is paid to the problem with the
commutativity of the fields' variations and the field operators or/and their
partial derivatives. It has been notice at first by Schwinger in his original
work~\cite{Schwinger-QFT-1} but later, in serious books
like~\cite{Roman-QFT}, it has been forgotten. A suitable solution of that
problem  is proposed in Sect.~\ref{Sect4} by giving a rigorous meaning of a
derivative of operator\ndash valued function of operator arguments with
respect to such an argument. It entails preservation of (operator)
Euler\ndash Lagrange equations for the field operators and a unique definition
of the operators of conserved quantities. A new moment is that the variations
of the field operators cannot be completely arbitrary in the general case (\eg
for some interacting fields) as they should satisfy some conditions derived in
this work. Sect~\ref{Sect5} illustrates the general theory of
Sect.~\ref{Sect4} with particular examples (free neutral or charged scalar
field, (self\ndash)interacting scalar fields, free spinor field, and system
of fields described via quadratic Lagrangian). It is presented an example of
a Lagrangian, describing free (or with some self\ndash interaction) spinor
field, for which the (classical operator) Euler\ndash Lagrange equations do
not exist in a sense that they are identities, like $0=0$. Regardless of that
fact, this Lagrangian entails completely reasonable field equations. The main
results of the work are summarized in Sect.~\ref{Conclusion}.

	In the Lagrangians we consider is not supposed normal ordering (of
the products of creation and annihilation operators). Besides, no
(anti)commu\-ta\-tion (or paracommutation) relations are supposed to be
fulfilled.  But the results obtained are, of course, valid and if normal
ordering of products is used some kind of (anti)commutation (or
paracommutation) relations are taken into account.


\section
[Problems with the equations of motion and with conserved\\ quantities]
{Problems with the equations of motion and\\ with conserved quantities}
\label{Sect2}

	Suppose a system of \emph{classical} fields $\varphi_i(x)$,
$i=1,\dots,n\in\field[N]$, over the Minkowski spacetime $M$, $x\in M$, is
described via a Lagrangian $L$ depending on them and their first partial
derivatives $\pd_\mu\varphi_i(x)=\frac{\pd\varphi_i(x)}{\pd x^\mu}$,
$\{x^\mu\}$ being the (local) coordinates of $x\in M$, \ie
 $L=L(\varphi_j(x), \pd_\nu\varphi_i(x))$. Here and henceforth the Greek
indices $\mu,\nu,\dots$ run from $0$ to $\dim M-1=3$ and the Latin indices
$i,j,\dots$ run from 1 to some integer $n$. The equations of motion for
$\varphi_i(x)$, known as the \emph{Euler\ndash Lagrange equations}, are%
\footnote{~%
In this paper the Einstein's summation convention over indices appearing twice
on different levels is assumed over the whole range of their values.%
}
	\begin{equation}	\label{2.1}
\frac{\pd L}{\pd \varphi_i(x)}
-
\frac{\pd}{\pd x^\mu} \Bigl( \frac{\pd L}{\pd (\pd_\mu\varphi_i(x))} \Bigr)
= 0
	\end{equation}
and are derived from the variational principle of stationary action, known
as the \emph{action principle} (see, e.g,~\cite[\S~1]{Bogolyubov&Shirkov},
\cite[\S~67]{Bjorken&Drell-2}, \cite[pp.~19\Ndash20]{Roman-QFT}).

	The (first) Noether theorem~\cite[\S~2]{Bogolyubov&Shirkov} says
that, if the action's variation is invariant under a $C^1$ transformations
	\begin{gather}
			\label{2.2new}
	\begin{split}
& x\mapsto x^\omega = x^\omega(x)
\quad x^\omega|_{\omega=\bs0} = x
\qquad \omega=(\omega^{(1)},\dots,\omega^{(s)})
\\
& \varphi_i(x) \mapsto \varphi_i^\omega(x^\omega)
\quad \varphi_i^\omega(x^\omega)|_{\omega=\bs0} = \varphi_i(x)
	\end{split}
\\\intertext{depending on $s\in\field[N]$ independent real parameters
$\omega^{(1)},\dots,\omega^{(s)}$, then the quantities}
			\label{2.2}
\theta_{(a)}^{\mu}(x)
:=
- \pi^{i\mu} \Bigl\{
\frac{\pd\varphi_i^\omega(x^\omega)} {\pd\omega^{(a)}} \Big|_{\omega=0}
 -
(\pd_\nu\varphi_i(x))
\frac{\pd x^{\omega\,\nu}} {\pd\omega^{(a)}} \Big|_{\omega=0}
	\Bigr\}
- L(x) \frac{\pd x^{\omega\,\mu}} {\pd\omega^{(a)}} \Big|_{\omega=0} ,
\intertext{where $a=1,\dots,s$ and}
			\label{2.3}
\pi^{i\mu} := \frac{\pd L}{\pd (\pd_\mu\varphi_i(x))} ,
\intertext{are conserved in a sense that}
			\label{2.4}
\pd_\mu \theta_{(a)}^{\mu}(x) = 0.
	\end{gather}
In particular, the invariance with respect to spacetime translations, \ie
$x\mapsto x^b=x+b$, with $b\in M$, and
$\varphi_i(x)\mapsto\varphi_i(x)$, leads to the conservation of the
\emph{energy\ndash momentum tensor}:
	\begin{align}	\label{2.5}
& T^{\mu\nu}(x):=\pi^{i\mu}(x) \pd^\nu\varphi_i(x) - L(x)\eta^{\mu\nu}
\\			\label{2.6}
& \pd_\nu T^{\mu\nu}(x) = 0,
	\end{align}
where $\eta^{\mu\nu}$ is the Lorentz metric tensor of $M$ with signature
$(+\,-\,-\,-)$ and the spacetime indices are raised (lowered) by
$\eta^{\mu\nu}$ (by the inverse tensor $\eta_{\mu\nu}$ of $\eta^{\mu\nu}$).
Analogously, the invariance relative to constant phase transformations,
\viz $x\mapsto x$ and
$\varphi_i(x)\mapsto \e^{\frac{q}{\ih c}\lambda}\varphi_i(x)$ where
$q=\const$, $\lambda$ is a real parameter, $\hbar$ is the Planck's constant
(divided by $2\pi$), and $c$ is the velocity of light in vacuum, implies the
conservation of the corresponding current:
	\begin{align}	\label{2.7}
& J_\mu(x)
:= \frac{q}{\ih c} \sum_{i} \varepsilon(\varphi_i) \pi^{i}_\mu(x) \varphi_i(x)
\\			\label{2.8}
& \pd^\mu J_\mu = 0
	\end{align}
where $\varepsilon(\varphi_i)=0$ if $\varphi_i(x)$ is real,
$\varepsilon(\varphi_i)=+1$ if $\varphi_i(x)$ is complex,  and
$\varepsilon(\varphi_j)=-1$ if $\varphi_j(x)$ is the complex conjugate
to $\varphi_i(x)$.%
\footnote{~%
It is a convention whether to a complex field or to its complex conjugate to
be assigned the value $+1$ of the function $\varepsilon$.%
}

	Below we shall be interested in the quantum case, when the fields
$\varphi_i(x)$ become linear operator depending on $x\in M$ and acting on
system's Hilbert space $\Hil$ of states. The above scheme is repeated
\emph{mutatis mutandis} in Heisenberg picture of motion in (canonical)
quantum field theory, when the fields $\varphi_i(x)$ are spacetime dependent
and the state vectors are spacetime independent. Details of this procedure
will be given in Sect.~\ref{Sect3} below. However, there are three related
problems which should find suitable answers:

	1.\ How the quantum Lagrangian $L$ should be defined? For example, if
a quantum system has a classical analogue, can we simply replace in the
classical Lagrangian the classical fields with the corresponding quantum
operators?

	2.\ What is the meaning of the derivative operators
$\frac{\pd}{\pd\varphi_i(x)}$ and
$\frac{\pd}{\pd(\pd_\mu\varphi_i(x))}$, appearing in, e.g., the Euler\ndash
Lagrange equations, when $\varphi_i(x)$ are operator\ndash valued, not
classical, functions?

	3.\ If the previous two problems are satisfactory  (well)  and
uniquely solved, how should be defined the conserved quantities~\eref{2.2} in
the quantum case? For instance, can we write the \emph{energy\ndash momentum
operator} as
	\begin{align}	\label{2.9}
T^{\mu\nu}(x):=\pi^{i\mu}(x)\circ( \pd^\nu\varphi_i(x)) - L(x)\eta^{\mu\nu} ?
	\end{align}
Here $\circ$ is the sign of mappings/operators composition/product and all
quantities are the operator analogues of the corresponding classical ones.

	Partially the nature of these problems is in the fact that,
generally, the field operators $\varphi_i(x)$ do not commute. So, the order in
which the field operators or functions of them appear in some composition
(product) is significant, contrary to the classical case.

	The solution of the first problem for the known fields, free or not,
has been found a long time
ago~\cite{Bogolyubov&Shirkov,Bjorken&Drell-2,Roman-QFT}. The basic
requirements for $L$ being that it should be a Hermitian operator which is
invariant under Lorentz/Poincar\'e transformations and other symmetries of
the system, if any.  Besides, if a quantum system has a classical analogue,
the classical Lagrangian should be equal to the quantum one when in the
latter the field operators are replaced with the corresponding classical
fields.

	\emph{A posteriori} there can be different solutions of the
second problem. However, \emph{a priori} there is a simple rule silently
followed in the literature~\cite{Bjorken&Drell-2, Roman-QFT}.  According to
it, one replaces in the quantum Lagrangian operator the field operators
$\varphi_i(x)$ (and their partial derivatives) with classical fields
$\varphi_i^{\text{cl}}(x)$ and the composition of mappings sign with
multiplication sign in such a way that the order of the (quantum) fields to be
preserved. Then, from the so\ndash obtained Lagrangian function
$L^{\text{cl}}$ are calculated the derivatives
 $\frac{\pd L^{\text{cl}}}{\pd\varphi_i^{\text{cl}}(x)}$ and
 $\frac{\pd L^{\text{cl}}}{\pd(\pd_\mu\varphi_i^{\text{cl}}(x))}$
by preserving the order of all fields and their derivatives. At the end, one
replaces in
 $\frac{\pd L^{\text{cl}}}{\pd\varphi_i^{\text{cl}}(x)}$ and
 $\frac{\pd L^{\text{cl}}}{\pd(\pd_\mu\varphi_i^{\text{cl}}(x))}$
the fields  $\varphi_i^{\text{cl}}(x)$ with the field operators
$\varphi_i(x)$ and the multiplication of fields with compositions of the
corresponding operators. In short, all this means that we differentiate a
quantum Lagrangian with respect to its operator arguments by the same rules
as in the classical case with the only addition that one should always retain
the initial order of all operators~\cite[\S~2]{Pauli&Heisenberg}. Following
this procedure, one should keep in mind that a change of the order of the
operators in the initial Lagrangian may result in different derivatives of it
even if the Lagrangian is not changed as an operator.%
\footnote{~%
If $A$ and $B$ are operators, the above rule implies
$\frac{\pd}{\pd A}(A\circ B) = \frac{\pd}{\pd A}(B\circ A) = B$.
So, if $A$ and $B$ anticommute, \ie $A\circ B=-B\circ A$, we have
$\frac{\pd}{\pd A}(A\circ B) = B $ and  $\frac{\pd}{\pd A}(-B\circ A)=-B$.
Consequently, the derivative relative to non\ndash commuting operator argument
has a `memory' for the place (left or right in our example) where the
arguments have been situated before the differentiation. This phenomenon will
find natural explanation in Sect.~\ref{Sect4}; in particular, see
remark~\ref{Rem4.1}.%
}

	When one analyzes the third of the afore presented problems, there
are two guiding principles: the conserved operators $\theta_{(a)}^{\mu}$
must be Hermitian and, if a system has a classical analogue, these operators
should reduce to the corresponding classical conserved fields~\eref{2.2} when
the field operators are replace with the corresponding to them classical
fields and the composition of operators is replaced by the multiplication of
(classical) fields. However, these guidelines are not enough for the explicit
determination of the conserved operators and one should `guess' their
functional form; the result can be justified or rejected \emph{a posteriori}
by examining the consequences of the model hypothesis.%
\footnote{~%
Since in quantum field theory are important the constant operators
$C_{(a)}:=\int_{\sigma}\theta_{(a)}^{\mu}\Id\sigma_\mu$, the integration
being along some 3\ndash dimensional spacelike surface $\sigma$, sometimes
different definitions of $\theta_{(a)}^{\mu}$ may result in identical
operators $C_{(a)}$, even if different Lagrangians are employed.%
}

	For instance, the straightforward transferring of~\eref{2.5} into the
quantum region results in~\eref{2.9} but this operator is, generally
non\ndash Hermitian. As a working hypothesis, one may assume a `Hermitian
symmetrization' of~\eref{2.9}, \viz
	\begin{equation}	\label{2.10}
T_{\mu\nu}
=
\frac{1}{2}\bigl\{
\pi_\mu^i(x)\circ(\pd_\nu\varphi_i(x))
	+ (\pd_\nu\varphi_i^\dag(x))\circ (\pi_\mu^i(x))^\dag
\bigr\}
- \eta_{\mu\nu} L(x)
	\end{equation}
where the dagger, ``$\dag$'', denotes Hermitian conjugation of operators. As
$L^\dag=L$, the last expression for energy\ndash momentum operator satisfies
the above\ndash written requirements.

	Similar is the situation with the current operator. \emph{Prima
facie} one may write (cf.~\eref{2.7})
	\begin{gather}	\label{2.11}
J_\mu
 = \frac{q}{\ih c}\sum_{i} \varepsilon(\varphi_i) \pi_\mu^i\circ\varphi_i
\\\intertext{but, generally, this expression is not Hermitian,
$J_\mu^\dag\not=J_\mu$. As a working hypothesis, a `Hermitian symmetrization'
may be assumed
(note, $\varepsilon(\varphi_i^\dag)=-\varepsilon(\varphi_i)$):}
			\label{2.12}
J_\mu
=
\frac{q}{\ih c}\sum_{i} \varepsilon(\varphi_i)\{
\pi_\mu^i\circ\varphi_i
-
\varphi_i^\dag\circ (\pi_\mu^i)^\dag	\} .
	\end{gather}

	A partial discussion of the above problems with
(energy\ndash)momentum and (current or) charge operator can be found
in~\cite{Pauli&Weisskopf}.


\section
[Schwinger's action principle (review and problems)]
{Schwinger's action principle (review and problems)}
\label{Sect3}

	The particular variant of the variation action principle, adapted to
the needs of quantum field theory, is known as the \emph{Schwinger's action
principle}. Its description can be found, for instance,
in~\cite[sec.~2.1]{Roman-QFT} or in the original paper~\cite{Schwinger-QFT-1}
(see also~\cite{Schwinger-QFT-2}). The purpose of the present section is a
concise summary of this method, some its consequences and problems it
contains. For details, the reader is referred to~\cite[sec.~2.1]{Roman-QFT},
from where the below\ndash presented resum\'e of Schwinger's action principle
is extracted.

	Let there be given a system of quantum fields represented via
linear field operators $\varphi_i(x)$. Let
$L=L(x)=L(\varphi_i(x),\pd_\mu\varphi_i(x))$ be the Lagrangian (density)
operator of the system. It is supposed to depend only on the field operators
and their first partial derivatives. Let $\sigma_1$ and $\sigma_2$ be two
3\ndash dimensional spacelike surfaces
and $R$ be the 4\ndash dimensional region (submanifold) bounded by them. The
\emph{action operator} is then defined by
	\begin{equation}	\label{3.1}
W
:= \frac{1}{c} \int\limits_{R} L(x) \Id^4x
=: \frac{1}{c}\int\limits_{\sigma_1}^{\sigma_2} L(x) \Id^4x  .
	\end{equation}
Consider the (infinitesimal) transformations
	\begin{subequations}	\label{3.2}
	\begin{align}	\label{3.2a}
x^\mu &\mapsto x^{\prime\,\mu} = x^\mu + \delta x^\mu
\\			\label{3.2b}
\varphi_i(x) &\mapsto \varphi'_i(x) + \delta_0\varphi_i(x)
	\end{align}
	\end{subequations}
as a result of a change of a spacetime point $x$ and field operator
$\varphi_i(x)$ when a transition to a \emph{new} reference frame is made;
in~\eref{3.2b} the symbol $x$ in $\varphi_i(x)$ refers to a point in the
\emph{new} frame. If $x\in\sigma$ for some spacelike surface $\sigma$,
it is supposed that the infinitesimal change $\delta_0\varphi_i(x)$ to be
generated by some generator $F[\sigma]$ which is operator\ndash valued
functional of $\sigma$, \ie
	\begin{equation}	\label{3.2new}
\delta_0\varphi_i(x) = \ih [F[\sigma],\varphi_i(x) ]_{\_}
	\end{equation}
where $[A,B]_{\pm}:=A\circ B \pm B\circ A$ for operators $A$ and $B$.

	The Schwinger's action principle postulates that, if~\eref{3.2}
induces the change $W\mapsto W+\delta W$ of the action integral~\eref{3.1},
then the infinitesimal change $\delta W$ of the action integral is a
difference of two surface integrals and
	\begin{equation}	\label{3.3}
\delta W = F[\sigma_2] - F[\sigma_1] .
	\end{equation}

	To work out consequences of~\eref{3.3}, we notice that the variation
$\delta W$ is due to independent effects of the variations
$\delta_0\varphi_i(x)$  of the field operators and the change $R\mapsto R'$
of the integration region as a result of the change~\eref{3.2a} of the points
of its boundary. So, neglecting second and higher order terms in the
variations and applying~\eref{3.1}, we can write
	\begin{gather}
			\label{3.4}
\delta W
=
\frac{1}{c}\int\limits_{R}
\Bigl\{ \delta_0 L + L \frac{\pd( \delta x^\mu )}{\pd x^\mu} \Bigr\} \Id^4x
\intertext{where}
			\label{3.5}
\delta_0 L
:=
L\bigl( \varphi_i(x) + \delta_0\varphi_i(x),
   \pd_\mu \varphi_i(x) + \pd_\mu(\delta_0\varphi_i(x)) \bigr)
-
L(\varphi_i(x),\pd_\mu\varphi_i(x))
\intertext{is the variation of the Lagrangian (operator).
\emph{Expanding the first term in~\eref{3.5} into Taylor series and neglecting
second and higher order terms, we get}}
			\label{3.6}
\delta_0 L
=
\frac{\pd L}{\pd\varphi_i(x)} \circ \delta_0\varphi_i(x)
+
\frac{\pd L}{\pd(\pd_\mu\varphi_i(x))} \circ \delta_0(\pd_\mu\varphi_i(x)) .
	\end{gather}
Here the derivatives are understood as described in Sect~\ref{Sect2}.

	\begin{Rem}	\label{Rem3.1}
	We have emphasized the last phrase because the transition
from~\eref{3.5} to~\eref{3.6} is generally incorrect and wrong, if
`arbitrary' variations $\delta_0\varphi_i(x)$ are considered. This will be
explained at length in Sect.~\ref{Sect4}. At this point, we recall only the
Schwinger's remark~\cite[the comments after eq.~(2.17)]{Schwinger-QFT-1} that
the expression~\eref{3.6} for $\delta_0 L$ should be considered as symbolic
one because it must be taken into account the (anti)commutation properties of
$\delta_0\varphi_i(x)$. To overcome this problem, he makes the hypothesis
that these properties and the structure of $L$ should be such that terms with
different positions of $\delta_0\varphi_i(x)$ must lead to equal portions in
the variation $\delta W$. As we shall see in Sect.~\ref{Sect4}, this is a
severe restriction which, generally, cuts off part of the information,
including the (anti)commutation properties of $\delta_0\varphi_i(x)$, the
Schwinger's action principle contains. The above problem, in other form, is
mentions in~\cite[p.~149]{Peierls} too.
	\end{Rem}

	\begin{Rem}	\label{Rem3.2}
	Alternatively, one can put the variations in~\eref{3.6} to the left
of the Lagrangian's derivatives. Such a modification does not change
anything, except the order of some operators, in the following. This problem
is partially mentioned in~\cite{Schwinger-QFT-2}.
	\end{Rem}

	Substituting~\eref{3.6} into~\eref{3.4}, noting that, by its
definition,
$\delta_0(\pd_\mu\varphi_i(x)) = \pd_\mu(\delta_0\varphi_i(x))$
and integrating by parts the term originating from the second term
in~\eref{3.6}, we obtain
	\begin{multline}	\label{3.7}
\delta W
=
\frac{1}{c}\int\limits_{R}\Bigl\{
\Bigl(
\frac{\pd L}{\pd\varphi_i(x)}
-
\frac{\pd}{\pd x^\mu} \Bigl( \frac{\pd L}{\pd(\pd_\mu\varphi_i(x))} \Bigr)
\Bigr) \circ \delta_0\varphi_i(x)
\\
+
\frac{\pd}{\pd x^\mu}\Bigl(
\frac{\pd L}{\pd(\pd_\mu\varphi_i(x))} \circ \delta_0\varphi_i(x)
	+ L \delta x^\mu\Bigr)
\Bigr\} \Id^4x \ .
	\end{multline}
Introducing the local (functional) variation
	\begin{equation}	\label{3.8}
\delta\varphi_i(x)
:= \varphi'_i(x') - \varphi_i(x)
 = \delta_0\varphi_i(x) + (\pd_\nu\varphi_i(x))\delta x^\nu,
	\end{equation}
where $x$ and $x'$ refer to the coordinates of one and the same geometric
point with respect to the `old' and `new' frames, applying the Stokes'
(Gauss') theorem, and repeating the steps in~\cite[p.~63]{Roman-QFT},
from~\eref{3.8} we get
	\begin{gather}	\label{3.9}
\delta W
=
\frac{1}{c} \int\limits_{R} \! \Bigl\{
\Bigl(
\frac{\pd L}{\pd\varphi_i(x)}
-
\frac{\pd}{\pd x^\mu} \Bigl( \frac{\pd L}{\pd(\pd_\mu\varphi_i(x))} \Bigr)
\Bigr) \circ \delta_0\varphi_i(x)
\Bigr\} \Id^4x \
+
F[\sigma_2] - F[\sigma_1]
	\end{gather}
where
	\begin{multline}
			\label{3.10}
F[\sigma]
: =
\frac{1}{c}\int\limits_{\sigma}\{
\pi^{i\mu}(x)\circ \delta_0\varphi_i(x) + L(x) \delta x^\mu
\} \Id\sigma_\mu
\\
 =
\frac{1}{c}\int\limits_{\sigma}\{
\pi^{i\mu}(x)\circ \delta \varphi_i(x)
-
\bigl( \pi^{i\mu}(x)\circ (\pd_\nu\varphi_i(x))
	- \delta_\nu^\mu L(x) \bigr) \delta x^\nu
\} \Id\sigma_\mu .
	\end{multline}
Here $\sigma$ is a spacelike surface with surface element $\Id\sigma_\mu$,
the notation~\eref{2.3} has been used, and $\delta_\mu^\nu$ is the (mixed)
Kroneker $\delta$\ndash symbol, \ie
$\delta_\mu^\nu=1$ for $\mu=\nu$ and
$\delta_\mu^\nu=0$ for $\mu\not=\nu$.

	Following the known argumentation (see~\cite[sec.~2.1]{Roman-QFT}
and~\cite[sec.~2]{Schwinger-QFT-1}), from~\eref{3.9} and~\eref{3.10} a number
of fundamental consequences can be derived. For example, we mention three of
them.

	Since~\eref{3.3} demands $\delta W$ to be a difference of two surface
integrals and $R$ and $\delta_0\varphi_i(x)$ are completely arbitrary,
from~\eref{3.9} the Euler\ndash Lagrange equations~\eref{2.1} for the field
operators $\varphi_i(x)$ follow.

	Identifying~\eref{3.10} with $\delta x^\nu=0$ with the generator of
the corresponding transformation~\eref{3.2} (with $\delta x^\nu=0$),
equation~\eref{3.2new} implies
	\begin{gather}
			\label{3.11}
\delta\varphi_i(x)
= \ih \Bigl[ \int\limits_{\sigma}
	\pi^{j\nu}(x')\circ \delta\varphi_j(x') \Id\sigma_\mu(x')
      , \varphi_i(x)
\Bigr]_{\_}
\intertext{as, by~\eref{3.8}, $\delta\varphi_i(x)=\delta_0\varphi_i(x)$ for
$\delta x^\mu=0$. Similarly}
			\label{3.12}
\delta\pi^{i\mu}(x)
= \ih \Bigl[ \int\limits_{\sigma}
	\pi^{j\nu}(x')\circ \delta\varphi_j(x') \Id\sigma_\mu(x')
      , \pi^{i\mu}(x)
\Bigr]_{\_} .
	\end{gather}
The last two equalities should be identities for arbitrary
$\delta\varphi_i(x)$ as long as
$(x'-x)^2:=(x^{\prime\,\mu}-x^\mu)(x^{\prime\,\nu}-x^\nu)\eta_{\mu\nu}<0$.
They can be satisfied if one assumes (as additional postulate) the famous
equal\ndash time (anti)commutation relations, as it is proved
in~\cite[pp.~65\Ndash67]{Roman-QFT}.

	Consider transformations~\eref{3.2} leaving the action operator
unchanged, \ie such that
	\begin{gather}
			\label{3.13}
\delta W = 0 .
\intertext{Then, by~\eref{3.9} and~\eref{2.1},}
			\label{3.14}
F[\sigma_1] = F[\sigma_2]
\intertext{for any 3-dimensional spacetime surfaces $\sigma_1$ and
$\sigma_2$, \ie the operators~\eref{3.10} are surface\ndash independent:}
\tag{\ref{3.14}$^\prime$}			\label{3.14'}
\frac{\delta F[\sigma]}{\delta\sigma(y)} = 0
\qquad y\in\sigma
\intertext{where $\frac{\delta}{\delta\sigma(y)}$ means the derivative of a
functional of $\sigma$ relative to $\sigma$ at
$y\in\sigma$~\cite[p.~10]{Roman-QFT}. From here follows that the `current
(density)'} 		\label{3.15}
f^\mu(x)
:=
\pi^{i\mu}(x)\circ \delta \varphi_i(x)
-
\bigl( \pi^{i\mu}(x)\circ (\pd_\nu\varphi_i(x))
	- \delta_\nu^\mu L(x) \bigr) \delta x^\nu
\intertext{is conserved, \viz}	\label{3.16}
\pd_\mu f^\mu(x) = 0.
	\end{gather}

It is a simple exercise to be verified, if we take~\eref{2.2new} as a
particular realization of~\eref{3.2}, then~\eref{3.15} and~\eref{3.16} will
respectively read
	\begin{gather}
			\label{3.17}
f^\mu(x) = - \sum_{a=1}^{s} \theta_{(a)}^{\mu}(x) \delta\omega^{(a)}
\\			 \label{3.18}
\pd_\mu  \theta_{(a)}^{\mu}(x) = 0 ,
\intertext{where}
			\label{3.19}
\theta_{(a)}^{\mu}(x)
:=
- \pi^{i\mu}\circ \Bigl\{
\frac{\pd\varphi_i^\omega(x^\omega)} {\pd\omega^{(a)}} \Big|_{\omega=0}
 -
(\pd_\nu\varphi_i(x))
\frac{\pd x^{\omega\,\nu}} {\pd\omega^{(a)}} \Big|_{\omega=0}
	\Bigr\}
- L(x) \frac{\pd x^{\omega\,\mu}} {\pd\omega^{(a)}} \Big|_{\omega=0}
	\end{gather}
is the quantum version of~\eref{2.2}. In this way, we arrive to the quantum
variant of the (first) Noether theorem saying that, if the action~\eref{3.1}
is invariant under finite parameter transformation~\eref{2.2new}, the
current operators~\eref{3.19} are conserved in a sense of~\eref{3.18} or,
equivalently, in a sense that the integrals
	\begin{gather}
			\label{3.20}
C_{(a)}(\sigma) := \int_\sigma \theta_{(a)}^{\mu} \Id\sigma_\mu
\intertext{are surface-independent, \ie}
			\label{3.21}
\frac{\delta C_{(a)}(\sigma)}{\delta\sigma(y)} = 0
\qquad y\in\sigma.
\\\intertext{ In particular, the choice $\sigma=\{x:x^0=ct=\const\}$ results
in}
\tag{\ref{3.20}$^\prime$}		\label{3.20'}
C_{(a)}(t) := \int_{x^0=ct} \theta_{(a)}^{\mu}(x) \Id^3\bs x
\\
\tag{\ref{3.21}$^\prime$}		\label{3.21'}
\frac{\od C_{(a)}(t)}{\od t} = 0 .
	\end{gather}

	We shall end the review of the Schwinger's action principle and its
consequences with the remark that the particular transformations~\eref{3.2}
with the choices
	\begin{align}	\label{3.22}
\delta x^\mu & = a^\mu	&& \delta_0\varphi_i(x) = 0
\\			\label{3.23}
\delta x^\mu & = 0	&& \delta_0\varphi_i(x) =
		\varepsilon(\varphi_i)\frac{q}{\ih c} \lambda \varphi_i(x) ,
	\end{align}
where $a^\mu$ and $\lambda$ are real parameters, lead to the canonical
energy\ndash momentum operator~\eref{2.9} and current operator~\eref{2.11},
respectively, and, consequently, to the accompanying them problems, as
discussed in Sect.~\ref{Sect2}.

	In remark~\ref{Rem3.1}, we mentioned that the
representation~\eref{3.6} for the r.h.s\ of~\eref{3.5} is, in the general
case, incorrect and wrong. Since the Lagrangian $L$ is supposed to be
polynomial or convergent power series in $\varphi_i(x)$ and
$\pd_\mu\varphi_i(x)$, it must be a sum of terms like
 $\alpha\psi_1(x)\circ\dots\circ\psi_a(x)$, where $\alpha$ is real or complex
number, $a\in\field[N]$ and $\psi_b(x)$, $b=1,\dots,a$, is a field operator
of a partial derivative of a field operator relative to some coordinate. The
variation of such a term, under the transformation~\eref{3.2b}, is
\[
\sum_{b=1}^{a}
  \psi_1(x)\circ\dots\circ\psi_{b-1}(x)
\circ\delta_0\psi_b(x) \circ
  \psi_{b+1}(x)\circ\dots\circ\psi_{a}(x)
\]
and can be put in the form~\eref{3.6}, \ie
\[
\Bigl(\sum_{b=1}^{a}
  \psi_1(x)\circ\dots\circ\psi_{b-1}(x)
\circ
  \psi_{b+1}(x)\circ\dots\circ\psi_{a}(x)
\Bigr) \circ \delta_0\psi_b(x),
\]
if and only if
	\begin{equation}	\label{3.24}
[ \delta_0\psi_b(x), \psi_{b+1}(x)\circ\dots\circ\psi_{a}(x) ]_{\_} = 0 .
	\end{equation}
These are the conditions Schwinger assumed to hold in~\cite[the comments
after eq.~(2.17)]{Schwinger-QFT-1} (see also~\cite{Schwinger-QFT-2}). The
particular form of these conditions depends, of course, on the concrete
Lagrangian under consideration. Generally, they say that the Lagrangian and
variations of the field operators cannot be completely independent and
arbitrary.

	Since, usually, the Lagrangian is considered as a basic object in the
theory, one may ask: can the conditions~\eref{3.24} hold for and arbitrary
Lagrangian? The answer is positive. For example, if the variations
$\delta_0\varphi_i(x)$  are chosen as multiples of the identity mapping
$\id_\Hil$ of the system's Hilbert space $\Hil$ of states, \ie
	\begin{gather}	\label{3.25}
\delta_0\varphi_i(x) = f_i(x) \id_\Hil
\\\intertext{ for \emph{completely arbitrary functions}
$f_i\colon M\to\field[C]$, then}
			\label{3.26}
\delta_0(\pd_\mu\varphi_i(x)) = (\pd_\mu f_i(x)) \id_\Hil
	\end{gather}
and, hence, the conditions~\eref{3.24} are identically satisfied. It is a
simple verification to be proved that the choices~\eref{3.25} are sufficient
for a rigorous derivation of all the results concerning Schwinger's action
principle reviewed above, as well as the ones in the
literature~\cite{Roman-QFT,Schwinger-QFT-1,Schwinger-QFT-2,Schwinger-QFT-3,
Peierls}.%
\footnote{~%
The choices~\eref{3.25} explain also the `usual' meaning of the derivatives
with respect to operators, as the ones in~\eref{2.1} and~\eref{3.7}.%
}
However, the choices~\eref{3.25} entail also the problems with the conserved
quantities mentioned in Sect.~\ref{Sect2}.

	As the purpose of the present paper is not the investigation of the
conditions under which~\eref{3.6} holds, we shall end this section with a
simple example illustrating the problem with~\eref{3.6}. Consider a free
neutral scalar field $\varphi(x)$ with mass parameter $m$. Its Lagrangian
is~\cite{Bjorken&Drell-2,Roman-QFT}
	\begin{gather}
			\label{3.27}
L
=
- \frac{1}{2}m^2c^4\varphi\circ\varphi
+ \frac{1}{2}c^2\hbar^2(\pd_\mu\varphi)\circ (\pd^\mu\varphi)
\\\intertext{so that}
			\label{3.28}
	\begin{split}
\delta L
=
 - \frac{1}{2}m^2c^4
	\{ \varphi\circ(\delta_0\varphi) + (\delta_0\varphi)\circ\varphi \}
+ \frac{1}{2}c^2\hbar^2\{
	(\pd_\mu\varphi)\circ (\delta_0(\pd^\mu\varphi))
	+(\delta_0(\pd_\mu\varphi))\circ (\pd^\mu\varphi) \} .
	\end{split}
	\end{gather}
Obviously, we can write the last expression in the form~\eref{3.6}, \ie as
	\begin{gather}
			\label{3.29}
\delta L
=
\frac{\pd L}{\pd\varphi} \circ \delta_0\varphi
+
\frac{\pd L}{\pd(\pd_\mu\varphi)} \circ \delta_0(\pd_\mu\varphi),
\\\intertext{where}
			\label{3.30}
\frac{\pd L}{\pd\varphi} = -m^2c^4\varphi
\quad
\frac{\pd L}{\pd(\pd_\mu\varphi)} = c^2\hbar^2 (\pd^\mu\varphi),
\intertext{if and only if}
			\label{3.31}
(\delta_0\varphi)\circ \varphi = \varphi\circ (\delta_0\varphi)
\quad
(\delta_0(\pd_\mu\varphi))\circ (\pd^\mu\varphi)
	=
	(\pd_\mu\varphi)\circ (\delta_0(\pd^\mu\varphi)) .
 	\end{gather}
If one assumes these equalities (or~\eref{3.29}) to hold for \emph{completely
arbitrary} $\delta_0\varphi$, as it is done everywhere in the literature,
(the Schur's lemma (see, e.g.,~\cite[sec.~8.2]{Kirillov-1976} or
\cite[ch.~5, sec.~3]{Barut&Roczka}) implies that) the field $\varphi$ must be
proportional to the identity mapping of system's Hilbert space $\Hil$,
$\varphi(x)=g(x)\id_\Hil$ with $g\colon M\to\field[C]$ of class $C^2$ such
that $m^2c^2g+\hbar^2(\pd_\mu\circ\pd^\mu)g=0$, due to~\eref{2.1} (\ie due to
the equation~\eref{4.2''} below). This is equivalent to a consideration of a
classical free real scalar field $g$. However, if we restrict the variety of
variations $\delta_0\varphi$ to the choice~\eref{3.25}, \ie
	\begin{equation}	\label{3.32}
\delta_0\varphi(x) = f(x) \id_\Hil
	\end{equation}
with completely arbitrary $f\colon M\to\field[C]$, we recover the standard
quantum field theory of a free neutral scalar field
$\varphi$~\cite{Roman-QFT,Bjorken&Drell-2,Bogolyubov&Shirkov}, accompanied
with the mentioned problems concerning the conserved quantities, \ie the
energy\ndash momentum tensorial operator in this particular case.


\section {A solution of the problems}
\label{Sect4}

	As we pointed in Sect.~\ref{Sect3}, a possible solution of the
problem with the representation~\eref{3.6} is to restrict the field
variations to multiples of the identity mapping $\id_\Hil$
(see~\eref{3.25}) which rigorously reproduces the known results and problems
following from Schwinger's action principle. However, it is our opinion, the
representation~\eref{3.6}, as well as all efforts to ensure its validity, is
not inherent to the Schwinger's variational principle of quantum field theory.
By imposing it, one restricts the possible Lagrangians and/or the variety of
possible variations of the field operators by a purely technical reason, which
is not in harmony with the other principles of quantum field theory. Moreover,
by demanding the validity of~\eref{3.6}, one `changes the rules of the game'
after it has been started, \viz after the variational principle is formulated
and the extraction of consequences of it has began, one suddenly imposes the
equality~\eref{3.6} only because it is valid in the classical case. We find
such a situation unsatisfactory and propose the below\ndash described
solution of all problems mentions until now. But, to illustrate the method we
intend to apply, we first consider the example Lagrangian~\eref{3.27}.

	Looking over~\eref{3.28} and~\eref{3.29}, we see that, for a free
neutral scalar field, the derivative $\frac{\pd L}{\pd\varphi}$ should be
regarded as a mapping acting on the operators on $\Hil$, not on vectors in
$\Hil$, such that
\(
\frac{\pd L}{\pd\varphi}\colon v \mapsto \frac{\pd L}{\pd\varphi}(v)
=
-\frac{1}{2}m^2c^4(\varphi\circ v + v\circ\varphi)
\)
for any $v\colon\Hil\to\Hil$. Defining similarly
$\frac{\pd L}{\pd(\pd_\mu\varphi)}$ by
\(
\frac{\pd L}{\pd(\pd_\mu\varphi)} \colon v
\mapsto
\frac{1}{2}c^2\hbar^2( (\pd^\mu\varphi)\circ v + v\circ(\pd^\mu\varphi) ),
\)
we can replace~\eref{3.29} with the equality
	\begin{gather}
			\label{4.1}
\delta L
=
\frac{\pd L}{\pd\varphi} ( \delta_0\varphi )
+
\frac{\pd L}{\pd(\pd_\mu\varphi)} ( \delta_0(\pd_\mu\varphi) )
	\end{gather}
where no additional conditions, like~\eref{3.31}, on $\varphi$ and
$\delta_0\varphi$ have been imposed. Further, repeating the derivation of
Euler\ndash Lagrange equations, with~\eref{4.1} for~\eref{3.6}, we get
	\begin{equation}	\label{4.2}
\Bigl(
\frac{\pd L}{\pd\varphi}
-
\frac{\pd}{\pd x^\mu} \Bigl(\frac{\pd L}{\pd(\pd_\mu\varphi)} \Bigr)
\Bigr)
(\delta_0\varphi)
= 0 .
	\end{equation}
Substituting here the just-obtained derivatives, we find
	\begin{gather}
\tag{\ref{4.2}$^\prime$}		\label{4.2'}
\frac{1}{2}(-m^2c^4\varphi - c^2\hbar^2 \square(\varphi) )
	\circ \delta_0\varphi
+
\delta_0\varphi \circ
	\frac{1}{2}(-m^2c^4\varphi - c^2\hbar^2 \square(\varphi) )
=0
\intertext{where $\square:=\pd_\mu\pd^\mu$ is the D'Alembert operator and no
additional condition have been imposed. The choice $\delta_0\varphi=\id_\Hil$
reduces the last equality to the standard Klein\ndash Gordon equation for
$\varphi$,}
\tag{\ref{4.2}$^{\prime\prime}$}	\label{4.2''}
m^2c^2\varphi + \hbar^2 \square(\varphi)  = 0
	\end{gather}
which, in turn, converts~\eref{4.2'} into identity. So, without the
\emph{additional} conditions~\eref{3.31}, we derived from the action
principle the `right'Klein\ndash Gordon equation describing free neutral
scalar field. For the energy\ndash momentum operator $T_{\mu\nu}$ of such a
field, one can repeat its derivation, as described in Sect.~\ref{Sect3}, but
with the new definition of the Lagrangian's derivatives. The result reads
(cf.~\eref{2.9})
	\begin{gather}
			 \label{4.3}
T_{\mu\nu} = \pi_\mu(\pd_\nu\varphi) - L \eta_{\mu\nu}
\intertext{where
\(
\pi^\mu := \frac{\pd L}{\pd(\pd_\mu\varphi)} \colon v
\mapsto
\frac{1}{2}c^2\hbar^2 ( (\pd^\mu\varphi)\circ v + v\circ (\pd^\mu\varphi) )
\)
for $v\colon\Hil\to\Hil$, or, equivalently}
\tag{\ref{4.3}$^\prime$}		\label{4.3'}
	\begin{split}
T_{\mu\nu}
=
 \frac{1}{2}c^2\hbar^2
	\{ (\pd_\mu\varphi)\circ (\pd_\nu\varphi)
	 + (\pd_\nu\varphi)\circ (\pd^\mu\varphi) \}
+
  \frac{1}{2}\{ m^2c^4\varphi\circ\varphi
	- c^2\hbar^2(\pd_\mu\varphi)\circ (\pd^\mu\varphi) \}
\eta_{\mu\nu} .
	\end{split}
	\end{gather}
Thus, without any additional hypotheses, the action principle~\eref{3.3}
leads to the `Hermitian symmetrized' energy\ndash momentum tensor~\eref{2.10}
(with the `usual' meaning of the $\pi^{i\mu}$, \ie
$\pi_\mu=c^2\hbar^2\pd_\mu\varphi$). It can be proved that, after imposing
the commutation relations and normal ordering, the quantum field $\varphi$
described via~\eref{4.2''} and~\eref{4.3'} is identical with the one
described via~\eref{4.2''} and
\(
T_{\mu\nu}
= c^2\hbar^2 (\pd_\mu\varphi)\circ (\pd_\nu\varphi)
	-  \eta_{\mu\nu} L,
\)
corresponding to~\eref{2.9}) and  usually considered in the
literature~\cite{Bogolyubov&Shirkov,Bjorken&Drell-2}.

	Having in mind the above example, we turn now our attention to the
general case of arbitrary Lagrangian, which is supposed to be polynomial or
convergent power series in the field operators and their first partial
derivatives.

	The main idea of the following is the derivatives of the Lagrangian
with respect to the field operators and/or their partial derivatives to be
defined as \emph{mappings acting on operators} such that (cf.~\eref{3.6}
and~\eref{4.1})
	\begin{equation}	\label{4.4}
\delta L
=
\sum_{i} \frac{\pd L}{\pd\varphi_i(x)} \bigl( \delta_0\varphi_i(x) \bigr)
+
\sum_{i,\mu}
\frac{\pd L}{\pd(\pd_\mu\varphi_i(x))}
	\bigl( \delta_0(\pd_\mu\varphi_i(x)) \bigr) .
	\end{equation}

	If one elaborates this definition, one will come the next definition
of a derivative of operator\ndash valued function of operator arguments with
respect to an operator variable.

	\begin{Defn}	\label{Defn4.1}
	Let $\Hil$ be a \field[C]-vector space,
$\omega\subseteq\{\Hil\to\Hil\}$ be a subset of the space of operators
acting on $\Hil$, $n\in\field[N]$,  $u_1,\dots,u_n\in\omega$,
$u\in\{u_1,\dots,u_n\}$, and
$f\colon(u_1,\dots,u_n)\mapsto f(u_1,\dots,u_n)\colon\Hil\to\Hil$ be
operator\ndash valued function of $u_1,\dots,u_n$ which is polynomial (or
convergent power series) in its operator arguments. The derivative of $f$
with respect to $u$ is an $n$\ndash argument mapping with domain
$\omega\times\dots\times\omega$ ($n$\ndash times),
denoted by $\frac{\pd f}{\pd u}$, such that:\\[0.17ex]
	\indent (i)
	Its value at $(u_1,\dots,u_n)$, denoted by
\(
\frac{\pd f}{\pd u} (u_1,\dots,u_n)
:=
\frac{\pd f(u_1,\dots,u_n)}{\pd u}
:=
\frac{\pd f}{\pd u}\big|_{(u_1,\dots,u_n)},
\)
is a mapping $\omega\to\{\Hil\to\Hil\}$ from the subset $\omega$ on the space
of operators on $\Hil$.\\[0.17ex]
	\indent (ii)
	The mapping $\frac{\pd}{\pd u}\colon f\mapsto \frac{\pd f}{\pd u}$
is linear relative to complex\ndash valued functions on $M$. In particular,
it is \field[C]\ndash linear.\\[0.17ex]
	\indent (iii)
	Let $v\colon\Hil\to\Hil$ be such that $u+v\in\omega$,
$a\in\field[N]$, $i_1,\dots,i_a\in\{1,\dots,n\}$ and
$I:=\{i\in\{i_1,\dots,i_a\} : u_i=u\}$ be the set of indices which label all
operators among $u_{i_1},\dots,u_{i_a}$ equal to $u$. Then
	\begin{equation}	\label{4.5}
\Bigl( \frac{\pd}{\pd u} \bigl(
	u_{i_1}\circ\dots\circ u_{i_a} \bigr)\Bigr) (v)
:=
\sum_{i\in I}
\Bigl\{\bigl( u_{i_1}\circ\dots\circ u_{i_a} \bigr) \big|_{u_i=v} \Bigr\} .
	\end{equation}
In particular, if $I$ is empty, $I=\varnothing$, the r.h.s.\ of~\eref{4.5} is
set equal to the zero operator of $\Hil$.
	\end{Defn}

	\begin{Rem}	\label{Rem4.1}
	The restriction $u+v\in\omega$ is essential one in quantum field
theory, in which the field operators, usually, satisfy some (anti)commutation
relations and, hence, in it $\omega\not=\{\Hil\to\Hil\}$; for some general
remarks on that item, see~\cite[sec.~21.1]{Ohnuki&Kamefuchi}. For instance,
let us find the derivative of $A\circ B$ with respect to $A$, where
$A,B\colon\Hil\to\Hil$ are anticommuting operators, $A\circ B=-B\circ A$. In
this particular case $\frac{\pd}{\pd A}(A\circ B)$ is defined only on those
$v\colon\Hil\to\Hil$ for which $(A+v)\circ B=-B\circ (A+v)$, \ie such that
$v\circ B=-B\circ v$, and hence
$\omega = \{ z\colon\Hil\to\Hil : z\circ B=-B\circ z \}$. In accord
with~\eref{4.5}, we have
\(
\frac{\pd(A\circ B)}{\pd A} (v)
= v\circ B = - B\circ v
=
\frac{\pd(- B\circ A)}{\pd A} (v);
\)
the evaluation of the derivative on element
$w\in\{\Hil\to\Hil\}\backslash\omega$ leads to a contradiction,
 $\frac{\pd(A\circ B)}{\pd A}(w) \not= \frac{\pd(-B\circ A)}{\pd A} (w)$.
	\end{Rem}

	\begin{Rem}	\label{Rem4.2}
	From a view-point of functional analysis, the
definition~\ref{Defn4.1} defines the notion of partial Fr\'echet derivative
of particular kind of functionals employed in quantum field theory. From this
position, the r.h.s.\ of~\eref{4.4} is nothing else, but the Fr\'echet
differential of the Lagrangian considered as a mapping between some operator
spaces.
       	\end{Rem}

	In short, definition~\ref{Defn4.1} means that a derivative of
operator-valued function, polynomial or convergent power series, of operator
arguments with respect to some of its operator arguments is calculated by
differentiating each its term according to~\eref{4.5}. In particular, this is
valid for Lagrangians of the type we consider in this work.

	It is a trivial checking to show that the derivatives introduce via
definition~\ref{Defn4.1} possess all `standard' derivative properties; in
particular, they satisfy the Leibnitz rule for differentiation of compositions
(products) of functions and the rule for differentiation of composite
functions.

	If $c\colon M\to\field[C]$ and $u+c(x)\id_\Hil\in\omega$, $x\in M$, it
is a trivial corollary of definition~\ref{Defn4.1} that
	\begin{equation}	\label{4.6}
\frac{\pd f(u_1,\dots,u_n)}{\pd u} (c(x)\id\Hil)
=
c(x) \frac{\pd^{\text{cl}} f(u_1,\dots,u_n)}{\pd u} ,
	\end{equation}
where all operators are supposed to be linear and $ \frac{\pd^{\text{cl}}
}{\pd u}$ means the `classical' derivative with respect to $u$ as it was
defined in Sect.~\ref{Sect2}, \ie the derivative in the r.h.s.\ of~\eref{4.6}
should be calculated as if $u_1,\dots,u_n$ were classical fields over $M$
with a preservation of the relative order of all operators. This is exactly
the definition of a derivative of a function of non\ndash commuting arguments
accepted, e.g., in~\cite[\S~2]{Pauli&Heisenberg}. For example, we have
\(
\frac{\pd\varphi^3}{\pd\varphi}(v)
=
\varphi^2\circ v + \varphi\circ v\circ\varphi + v\circ\varphi^2
\)
and
\(
\frac{\pd\varphi^3}{\pd\varphi}(c(x)\id_\Hil) = 3c(x)\varphi^2
\)
where $\varphi^a:=\varphi\circ\dots\varphi$ ($a$\ndash times) for
$a\in\field[N]$  and $\varphi\colon\Hil\to\Hil$.

	Equipped with the new definition of a derivative of a Lagrangian with
respect to a field operator or its partial derivative, it is trivial to verify
that~\eref{4.4} is an equivalent version of~\eref{3.5}, up to second and
higher order terms, without making any additional hypotheses with respect to
the Lagrangian or/and variations of the field operators. Moreover, if we
consider the variations~\eref{3.25}, then, in view of~\eref{4.6}
and~\eref{4.4}, we get
	\begin{multline}	\label{4.7}
\delta_0 L|_{\delta_0\varphi(x)=f_i(x)\id_\Hil}
=
\Bigl\{
\frac{\pd^{\text{cl}} L}{\pd\varphi_i(x)} \circ \delta_0\varphi_i(x)
+
\frac{\pd^{\text{cl}} L}{\pd(\pd_\mu\varphi_i(x))}
	\circ \delta_0(\pd_\mu\varphi_i(x))
\Bigr\}\Big|_{\delta_0\varphi(x)=f_i(x)\id_\Hil}
\\
=
f_i(x) \frac{\pd^{\text{cl}} L}{\pd\varphi_i(x)}
+
(\pd_\mu f_i(x)) \frac{\pd^{\text{cl}} L}{\pd(\pd_\mu\varphi_i(x))}
	\end{multline}
from where all standard consequences (and problems) of the Schwinger's action
principle can be derived (recovered).%
\footnote{~%
See the remark in~\cite[p.~149]{Peierls} on the above topic.%
}
It is natural to be expected that the restriction of the field variations to
ones given via~\eref{3.25} should lead to a broadening of the consequences of
the variational principle~\eref{3.3}. Below we shall examine them without
accepting any additional conditions, like~\eref{3.25} or, in the general
case,~\eref{3.24}. In this case, one can expect new consequences of the
Schwinger's action principle, which otherwise are `swallowed' by~\eref{3.25}
(or~\eref{3.24}) in its standard presentations.

	We shall now work out the explicit form of the action
variation~\eref{3.4} with~\eref{3.6} replaced by~\eref{4.4} with the new
meaning of the derivatives in it. Inserting~\eref{4.4} into~\eref{3.4} and
integrating by parts the term coming from the second one in~\eref{4.4},%
\footnote{~%
By virtue of properties (ii) and (iii) in definition~\ref{Defn4.1}, the
integration by parts is a rigorous operation; in general, we have:
\(
\frac{\pd f}{\pd u}(\pd_\mu v)
=
\pd_\mu\bigl(\frac{\pd f}{\pd u} (v) \bigr)
	- \bigl(\pd_\mu\bigl(\frac{\pd f}{\pd u}\bigr)\bigr)(v).
\)%
}
we get (cf~\eref{3.7})
	\begin{multline}	\label{4.8}
\delta W
 =
\frac{1}{c}\int\limits_{R}
\Bigl\{ \sum_{i}
\Bigl(
\frac{\pd L}{\pd\varphi_i(x)}
- \frac{\pd}{\pd x^\mu}
	\Bigl( \frac{\pd L}{\pd(\pd_\mu\varphi_i(x))} \Bigr)
\Bigr) \bigl( \delta_0\varphi_i(x) \bigr)
\\
+
\sum_{\mu} \frac{\pd}{\pd x^\mu}
\Bigl( \sum_{i}
\frac{\pd L}{\pd(\pd_\mu\varphi_i(x))}
	\bigl( \delta_0 \varphi_i(x) \bigr)
+ L(x) \delta x^\mu
\Bigr)
\Bigr\} \Id ^4x .
	\end{multline}
From here, repeating \emph{mutatis mutandis} the derivation of~\eref{3.9}, as
given in~\cite{Roman-QFT}, we obtain
	\begin{gather}
			\label{4.9}
	\begin{split}
\delta W
 =
\frac{1}{c}\int\limits_{R}
\Bigl\{ \sum_{i}
\Bigl(
\frac{\pd L}{\pd\varphi_i(x)}
- \frac{\pd}{\pd x^\mu}
	\Bigl( \frac{\pd L}{\pd(\pd_\mu\varphi_i(x))} \Bigr)
\Bigr) \bigl( \delta_0\varphi_i(x) \bigr)
\Bigr\} \Id ^4x
+
F[\sigma_2] - F[\sigma_1]
	\end{split}
\\\intertext{where}
			\label{4.10}
	\begin{split}
F[\sigma]
:=
\frac{1}{c} &\int\limits_{\sigma} \sum_{\mu}
\Bigl\{\sum_{i}
\pi^{i\mu}(x) ( \delta\varphi_i(x) )
 -
\sum_{\nu}\Bigl( \sum_{i} \pi^{i\mu}(x)
	(\pd_\nu\varphi_i(x)) -\delta_\nu^\mu L(x) \Bigr) \delta x^\nu
\Bigr\} \Id\sigma_\mu
	\end{split}
\\\intertext{with}
			\label{4.11}
\pi^{i\mu}(x) := \frac{\pd L}{\pd(\pd_\mu\varphi_i(x))}
\colon
\{\Hil\to\Hil\} \to \{\Hil\to\Hil\}
	\end{gather}
being the derivative of $L$ relative to $\pd_\mu\varphi_i(x)$ according to
definition~\ref{Defn4.1}. Formally,~\eref{4.9} and~\eref{4.10} can be
obtained from~\eref{3.9} and~\eref{3.10} via the next replacements:
	\begin{equation}	\label{4.12}
	\begin{split}
&
\frac{\pd L}{\pd\varphi_i(x)} \circ v
\mapsto
\frac{\pd L}{\pd\varphi_i(x)} \bigl( v\bigr)
\qquad
\Bigl\{
\frac{\pd}{\pd x^\mu}\Bigl( \frac{\pd L}{\pd\varphi_i(x)} \Bigr)
\Bigr\} \circ v
\mapsto
\Bigl\{\frac{\pd}{\pd x^\mu}
	\Bigl( \frac{\pd L}{\pd\varphi_i(x)} \Bigr) \Bigr\} \bigl( v\bigr)
\\ &
\pi^{i\mu}\circ v = \frac{\pd L}{\pd(\pd_\mu\varphi_i(x))} \circ v
\mapsto
\pi^{i\mu} (v ) = \frac{\pd L}{\pd(\pd_\mu\varphi_i(x))} \bigl( v\bigr)
	\end{split}
	\end{equation}
where $v\colon\Hil\to\Hil$ (is some variation, \ie $\delta_0\varphi_i(x)$ or
$\delta_0\varphi_i$). As we shall see further, these changes can be used for
`repairing' the standard consequences of Schwinger's action principle.
Evidently, the replacements opposite to~\eref{4.12} transform~\eref{4.9}
and~\eref{4.10} to~\eref{3.9} and~\eref{3.10}, respectively.

	Evidently, in view of~\eref{4.6}, it is clear that the `old'
variation~\eref{3.9} (with the standard meaning of the derivatives in it)
and the `new' variation~\eref{4.9} (with the derivatives in it given via
definition~\ref{Defn4.1}) are identical if variations like~\eref{3.25}, \ie
multipliers of the identity mapping, are employed.

	Let us proceed with extraction of consequences of the variational
principle~\eref{3.3} on a base of the representation~\eref{4.9}.

	Since~\eref{3.3} states that  $\delta W$ must be a difference of two
surface integrals,~\eref{4.9} implies the vanishment of the volume integral
in it, which, due to the arbitrariness of the integration region $R$, is
equivalent to
	\begin{equation}	\label{4.13}
\sum_{i}\Bigl\{
\Bigl(
\frac{\pd L}{\pd\varphi_i(x)}
- \frac{\pd}{\pd x^\mu}
	\Bigl( \frac{\pd L}{\pd(\pd_\mu\varphi_i(x))} \Bigr)
\Bigr) \bigl( v_i \bigr)
\Bigr\} = 0
	\end{equation}
where, for brevity, we have denoted by $v_i$ the variation
$\delta_0\varphi_i(x)$  of $\varphi_i(x)$, $v_i:=\delta_0\varphi_i(x)$. This
is the prototype of the \emph{operator Euler\ndash Lagrange equations for the
field operators $\varphi_i(x)$ and their variations
$v_i:=\delta_0\varphi_i(x)$}. We should emphasize, now this is an equation
both for $\varphi_i(x)$ and $v_i$, contrary to the standard procedure where
one gets, due to the arbitrariness of $v_i$, equations only for
$\varphi_i(x)$.%
\footnote{~%
This does not exclude the coincidence of the final equations for $\varphi_i$;
in particular, such is the case with the free fields --- see
Sect.~\ref{Sect5}.%
}
So, the new moment with respect to the `old' variational principle is
that~\eref{4.13} puts, in general, restrictions both on the field operators
and on their variations, \ie the variations cannot be considered as
completely arbitrary (if one does not wont to deal with trivial fields in
some cases). In Sect.~\ref{Sect5}, examples will be considered when
completely arbitrary and not such variations are admissible; the particular
situation depends on the concrete Lagrangian employed.%
\footnote{~%
As a rule, the Lagrangians describing free fields admit arbitrary variations,
while those describing (self\ndash)interacting fields require some
restrictions on the variety of field variations.%
}
However, from the derivation of~\eref{4.9} (and therefore of~\eref{4.13}), it
is clear that the variations~\eref{3.25}, \ie ones proportional to the
identity mapping $\id_\Hil$ of the system's Hilbert space $\Hil$ of states,
are always admissible. For them, in view of~\eref{4.6} and the complete
arbitrariness of the functions $f_i\colon M\to\field[C]$ in~\eref{3.25}, we
derive from~\eref{4.13} the `classical' Euler\ndash Lagrange equations for
the field operators as
	\begin{equation}	\label{4.14}
\frac{\pd^{\text{cl}} L}{\pd\varphi_i(x)}
- \frac{\pd}{\pd x^\mu}
	\Bigl( \frac{\pd^{\text{cl}} L}{\pd(\pd_\mu\varphi_i(x))} \Bigr)
= 0
	\end{equation}
which, due to the meaning of the operator derivatives in it, coincide with
the ones obtained form the `old' variations~\eref{3.9}. But it should clearly
be understood, if one requires~\eref{4.13} to be valid for \emph{completely
arbitrary} variations $v_i$, other restrictions on the field operators
may arise.%
\footnote{~%
In some cases, these new restrictions imply the field operators to be
proportional to $\id_\Hil$, i.e., in a sense, leading to classical, not
quantum, fields.%
}
The alternative point of view is to look on~\eref{4.14} as
on field equations for the field operators and on the remaining  consequences
of~\eref{4.13}, if any, as on restrictions on the admissible variations $v_i$.
We shall discuss this topic in Sect.~\ref{Sect5} on concrete examples; in
particular, in Subsect.~\ref{Subsect5.5new} it will be presented an example
of a Lagrangian which leads to completely reasonable field equations which
are \emph{not} the Euler\ndash Lagrange equations for it (the latter being
simply identities with respect to the fields and their variations).

	Let us turn now to the problem with conserved quantities. Consider a
transformation~\eref{3.2}, in which the field operators $\varphi_i(x)$ and
their variations $v_i=\delta_0\varphi_i(x)$ satisfy~\eref{4.13}, leaving the
action~\eref{3.1} unchanged. For these operators
equations~\eref{3.1}--\eref{3.14'} hold (see~\eref{4.9} and~\eref{4.13}) with
$F[\sigma]$ defined, now, by~\eref{4.10}, not by~\eref{3.10}. Therefore, the
`current(density)' (cf.~\eref{3.15})
	\begin{equation}	\label{4.15}
f^\mu(x)
:=
\sum_{i} \pi^{i\mu}(x) ( \delta\varphi_i(x) )
-
\sum_{\nu}\Bigl( \sum_{i}  \pi^{i\mu}(x)
	(\pd_\nu\varphi_i(x)) -\delta_\nu^\mu L(x) \Bigr) \delta x^\nu
	\end{equation}
is conserved, \ie satisfies the continuity equation~\eref{3.16}. In
particular, if an $s$\ndash parameter transformations~\eref{2.2new} satisfy
the above conditions, the equalities~\eref{4.15} and~\eref{3.16} will reduce
respectively to~\eref{3.17} and~\eref{3.18} with the `Noether currents'
$\theta_{(a)}^{\mu}(x)$, $a=1,\dots,s$, given by (cf~\eref{3.19})
	\begin{equation}	\label{4.16}
\theta_{(a)}^{\mu}(x)
:=
- \sum_{i} \pi^{i\mu}(x)
\Bigl(
\frac{\pd\varphi_i^\omega(x^\omega)} {\pd\omega^{(a)}} \Big|_{\omega=0}
\Bigr)
+
\sum_{i,\nu}\pi^{i\mu}(x)
\bigl( \pd_\nu\varphi_i(x) \bigr)
\frac{\pd x^{\omega\,\nu}} {\pd\omega^{(a)}} \Big|_{\omega=0}
-
L(x) \frac{\pd x^{\omega\,\mu}} {\pd\omega^{(a)}} \Big|_{\omega=0}
	\end{equation}
where $\pi^{i\mu(x)}$ is defined via~\eref{4.11}. The
equations~\eref{3.20}--\eref{3.21'}, of course, remain valid with the new
definition~\eref{4.16} of $\theta_{(a)}^{\mu}(x)$.

	To feel better the difference between~\eref{4.15} and~\eref{3.15} (or
between~\eref{4.16} and~\eref{3.19}), let us consider the
transformations~\eref{3.22} and~\eref{3.23}, generating in the classical case
the energy\ndash momentum tensor~\eref{2.5} and current vector~\eref{2.7},
respectively. For them the operator~\eref{4.15} reduces respectively to
 $f^\mu(x)=-\Sprindex[T]{\nu}{\mu}(x) a^\nu$ and
 $f^\mu(x) = J^\mu(x) \lambda$, where
	\begin{align}	\label{4.17}
T^{\mu\nu}(x)
&=
\sum_{i} \pi^{i\mu}(x)\bigl(\pd^\nu\varphi_i(x)\bigr) -\eta^{\mu\nu} L(x)
\\
			\label{4.18}
J^\mu(x)
&=
\frac{q}{\ih c} \sum_{i}
	\varepsilon(\varphi_i) \pi^{i\mu}(x)\bigl(\varphi_i(x)\bigr)
	\end{align}
are the energy\ndash momentum and (charge) current operators, respectively.
In some cases, these expressions may differ significantly from~\eref{2.9}
and~\eref{2.11}, respectively, which will be illustrated on concrete examples
in Sect.~\ref{Sect5}.

	As a last example of a conserved operator quantity, we consider the
angular momentum operator $M_{\mu\nu}^{\lambda}$. Suppose the action operator
is invariant under 4\ndash rotations, \ie under the changes
$x^\mu\mapsto x^{\varepsilon\,\mu}=x^\mu+\varepsilon^{\mu\nu} x_\nu$,
with $x_\nu:=\eta_{\nu\mu}x^\mu$ and antisymmetric real parameters
$\varepsilon^{\mu\nu}=-\varepsilon^{\nu\mu}$, and
 $\varphi_i(x)\mapsto\varphi_i^\varepsilon(x^\varepsilon)$ with
\(
\varphi_i^\varepsilon(x^\varepsilon)
=
\varphi_i(x)
+ \sum_{\mu<\nu}I_{i\mu\nu}^{j} \varphi_j(x) \varepsilon^{\mu\nu}
+\dotsb ,
\)
where the dots stand for second and higher order terms in
$\varepsilon^{\mu\nu}$ and $I_{i\mu\nu}^{j}=-I_{i\nu\mu}^{j}$ are numbers
characterizing the behaviour of the field operators under 4\ndash rotations.
Since
\(
x^{\varepsilon\,\rho}
=
x^\rho + \sum_{\mu<\nu}(\delta_\mu^\rho x_\nu - \delta_\nu^\rho x_\mu)
	\varepsilon^{\mu\nu} ,
\)
from~\eref{4.16} (with changed sign), in view of~\eref{4.17}, we obtain
	\begin{gather}	\label{4.19}
M_{\mu\nu}^{\lambda}
=
\bigl( x_\mu \Sprindex[T]{\nu}{\lambda} - x_\nu \Sprindex[T]{\mu}{\lambda}
\bigr)
+ S_{\mu\nu}^{\lambda} ,
\intertext{where}
				\label{4.20}
S_{\mu\nu}^{\lambda}
:=
\sum_{i,j} \pi^{i\lambda}(\varphi_j) I_{i\mu\nu}^{j}
	\end{gather}
is the spin angular momentum operator.

	As for the quantization rules, equation~\eref{3.11} should now be
replaced by
	\begin{equation}	\label{4.21}
\delta\varphi_i(x)
=
\ih\int_{\sigma}
\Bigl[\sum_{j} \pi^{j\nu}(x') \bigl(\delta\varphi_j(x')\bigr)
,
\varphi_i(x) \Bigr]_{\_} \Id\sigma_\nu(x')
	\end{equation}
and similarly for~\eref{3.12}. These equations and the
variations~\eref{3.25}, combined with the know
argumentation~\cite[sect~2.1~(ii)]{Roman-QFT}, produce the canonical
(anti)commutation relations. However, a different choice of the field
variations, if such ones are admissible, may result in new restrictions on
the field operators.


\section {Examples}
\label{Sect5}

	The main purpose of this section is an illustration of the general
theory of Sect.~\ref{Sect4} for particular Lagrangians. As we shall wee,
known results are reproduce with some corrections. The `quadratic'
Lagrangians will be pointed as the `best' ones selected by the Schwinger's
action principle.

\subsection{Free neutral scalar field}
\label{Subsect5.1}

	The Lagrangian of a free neutral scalar field $\varphi=\varphi^\dag$
with mass parameter $m$ is~\eref{3.27}. In accord with
definition~\ref{Defn4.1}, the action of its operator derivatives on an operator
$v$ are (cf.~\eref{3.30})
	\begin{gather}
				\label{5.1}
	\begin{split}
& \frac{\pd L}{\pd\varphi} (v)
=
- \frac{1}{2}m^2c^4( \varphi\circ v + v\circ\varphi )
\\ &
\pi^\mu(v)= \frac{\pd L}{\pd(\pd_\mu\varphi)} (v)
=
\frac{1}{2}c^2\hbar^2
    \bigl( (\pd^\mu\varphi)\circ v + v\circ(\pd^\mu\varphi) \bigr) .
	\end{split}
\intertext{Hence the Euler-Lagrange relation~\eref{4.13} for it reads}
				\label{5.2}
\frac{1}{2}( -m^2c^4\varphi - c^2\hbar^2 \square(\varphi) ) \circ v
+
v\circ \frac{1}{2}( -m^2c^4\varphi - c^2\hbar^2 \square(\varphi) )
= 0
\intertext{where $\square=\pd_\mu\pd^\mu$ is the D'Alembert operator. The
choice $v=\id_\Hil$,  $\Hil$ being the field's (system's) Hilbert space of
states, results in the Klein\ndash Gordon equation}
				\label{5.3}
m^2c^2\varphi + \hbar^2\square(\varphi) = 0
	\end{gather}
which corresponds to~\eref{4.14} with Lagrangian~\eref{3.27}.
Evidently,~\eref{5.3} converts~\eref{5.2} into identity relative to $v$. Thus
the equations~\eref{4.14} do not impose any restrictions for the variations
$v$ in a case of the Lagrangian~\eref{3.27}. In view of~\eref{5.1}, the
energy\ndash momentum operator~\eref{4.17} now reads
	\begin{equation}	\label{5.4}
T_{\mu\nu}
=
\frac{1}{2}c^2\hbar^2
\bigl\{ (\pd_\mu\varphi)\circ(\pd_\nu\varphi) +
			(\pd_\nu\varphi)\circ(\pd_\mu\varphi) \bigr\}
- \eta_{\mu\nu} \bigl\{ - \frac{1}{2}m^2c^4\varphi\circ\varphi +
\frac{1}{2}c^2\hbar^2(\pd_\mu\varphi)\circ (\pd^\mu\varphi) \bigr\} ,
	\end{equation}
which corresponds to~\eref{2.10} with Lagrangian~\eref{3.27}, not
to~\eref{2.9}, when the additional conditions~\eref{3.24}, i.e.~\eref{3.31} in
the particular case, in Schwinger's action principle are imposed.

	In almost the same way, the reader may wish to consider an
electromagnetic field with 4\ndash potential operators $A_\mu$ and, e.g.,
gauge invariant Lagrangian
$L=-\frac{1}{4}c^2\hbar^2F_{\mu\nu}F^{\mu\nu}$ with
$F_{\mu\nu}:=\pd_\mu A_\nu - \pd_\nu A_\mu$.
In this case, we have
$\frac{\pd L}{A^\mu} = 0$ and (do not sum over $\mu$!)
\(
\pi_\mu^\nu(v_\mu)
= \frac{\pd L}{\pd(\pd_\nu A^\mu)} (v_\mu)
= \frac{1}{2}c^2\hbar^2 \{ v_\mu \circ F_{\nu\mu} + F_{\nu\mu} \circ v_\mu \}
\)
for a variation $v_\mu$ of $A_\mu$.

\subsection{Free charged scalar field}
\label{Subsect5.2}

	The standard choice of a Lagrangian of a free charged scalar field
$\varphi\not=\varphi^\dag$ with mass parameter $m$
is~\cite{Bogolyubov&Shirkov,Bjorken&Drell-2,Roman-QFT}%
\footnote{~%
In some sense, the Lagrangian
\(
L =
- \frac{1}{2}m^2c^4 (\varphi^\dag\circ\varphi+\varphi\circ\varphi^\dag)
+ \frac{1}{2}c^2\hbar^2
  \bigl( (\pd_\mu\varphi^\dag)\circ(\pd^\mu\varphi)
	+ (\pd_\mu\varphi)\circ(\pd^\mu\varphi^\dag) \bigr)
\)
is better than~\eref{5.5}. But, after imposing the commutation relations
and normal ordering, the quantum field theory arising from both Lagrangians
turns to be one and the same.%
}
	\begin{gather}
			\label{5.5}
L =
- m^2c^4 \varphi^\dag\circ\varphi
+ c^2\hbar^2  (\pd_\mu\varphi^\dag)\circ(\pd^\mu\varphi) .
\\\intertext{So, we have:}
			\label{5.6}
	\begin{split}
\frac{\pd L}{\pd\varphi} (v) = -m^2c^4 \varphi^\dag\circ v
\quad
\pi^\mu(v) = \frac{\pd L}{\pd(\pd_\mu\varphi)} (v)
=
c^2\hbar^2(\pd^\mu\varphi^\dag)\circ v
\\
\frac{\pd L}{\pd\varphi^\dag} (w) = -m^2c^4 w\circ\varphi\circ
\quad
\pi^{\dag\,\mu}(w) = \frac{\pd L}{\pd(\pd_\mu\varphi^\dag)} (w)
=
c^2\hbar^2 w\circ(\pd^\mu\varphi)
	\end{split}
	\end{gather}
for operators $v$ and $w$ having a meaning of variations of $\varphi$ and
$\varphi^\dag$, respectively. Therefore the relation~\eref{4.13} now reads:
	\begin{equation}
			\label{5.7}
(-m^2c^4\varphi^\dag -c^2\hbar^2\square(\varphi^\dag) ) \circ v
+
w \circ (-m^2c^4\varphi - c^2\hbar^2\square(\varphi) ) = 0 .
	\end{equation}
From here, using the standard choices $(v,w)=(0,\id_\Hil),(\id_\Hil,0)$, we
derive the Klein\ndash Gordon equations
	\begin{equation}
			\label{5.8}
m^2c^2\varphi + \hbar^2\square(\varphi ) = 0
\qquad
m^2c^2\varphi^\dag + \hbar^2\square(\varphi^\dag) = 0 ,
	\end{equation}
which convert~\eref{5.7} into identity relative to $v$ and $w$. Thus, the
variations $v$ and $w$ in~\eref{5.7} can be completely arbitrary.
Substituting~\eref{5.6} into~\eref{4.17} and~\eref{4.18}, we get the
energy\ndash momentum and current operators respectively as
	\begin{gather}	\label{5.9}
	\begin{split}
T_{\mu\nu}
=
c^2\hbar^2
\bigl\{ (\pd_\mu\varphi^\dag)
 \circ(\pd_\nu\varphi) + (\pd_\nu\varphi^\dag)\circ(\pd_\mu\varphi) \bigr\}
- \eta_{\mu\nu} \bigl\{
- m^2c^4 \varphi^\dag\circ\varphi
+ c^2\hbar^2  (\pd_\mu\varphi^\dag)\circ(\pd^\mu\varphi)
 \bigr\}
	\end{split}
\\			\label{5.10} \mspace{-209mu}
J_\mu
=
-\ih cq \bigl\{
(\pd_\mu\varphi^\dag)\circ\varphi - \varphi^\dag\circ(\pd_\mu\varphi)
\bigr\} ,
	\end{gather}
where, for definiteness, we have chosen $\varepsilon(\varphi)=+1$ and
$\varepsilon(\varphi^\dag)=-1$. These expressions correspond to~\eref{2.10}
and~\eref{2.12}, not to~\eref{2.9} and~\eref{2.11}, respectively. Evidently,
$T_{\mu\nu}=T_{\nu\mu}$, $T_{\mu\nu}^\dag=T_{\mu\nu}$, and
$J_\mu^\dag=J_\mu$. Consequently, our formalism produces the known results
form the literature~\cite{Bogolyubov&Shirkov,Bjorken&Drell-2}, where
expressions, like~\eref{5.9} and~\eref{5.10}, are more a matter of
convention/postulate than a one of rigorous derivation.

\subsection{Self-interacting neutral scalar field}
\label{Subsect5.3}

	Consider a neutral scalar field $\varphi=\varphi^\dag$ with
Lagrangian
	\begin{equation}	\label{5.11}
L
=
- \frac{1}{2}m^2c^4\varphi\circ\varphi
+ \frac{1}{2}c^2\hbar^2(\pd_\mu\varphi)\circ (\pd^\mu\varphi)
+ a \varphi^k
	\end{equation}
where $a$ is a real non-zero parameter, $k\in\field[N]$, and
$\varphi^k:=\varphi\circ\dots\varphi$ ($k$\ndash times). Applying
definition~\ref{Defn4.1}, we get:
	\begin{gather}
			\label{5.12}
	\begin{split}
\frac{\pd L}{\pd\varphi} (v)
& =
- \frac{1}{2}m^2c^4 (\varphi\circ v + v\circ\varphi)
\\&
+ a
( v\circ\varphi^{k-1} + \varphi\circ v\circ\varphi^{k-2} + \dots
	+ \varphi^{k-2}\circ v\circ\varphi + \varphi^{k-1}\circ v )
\\
\pi^\mu(v) &= \frac{\pd L}{\pd(\pd_\mu\varphi)} (v)
=
\frac{1}{2}c^2\hbar^2
	\bigl( (\pd_\mu\varphi)\circ v + v\circ(\pd_\mu\varphi) \bigr) .
	\end{split}
	\end{gather}
So, the energy\ndash momentum operator is given by~\eref{5.4}, with the term
in the braces replaced by the r.h.s.\ of~\eref{5.11}, and the
relation~\eref{4.13} reads (cf.~\eref{5.2})
	\begin{multline}
			\label{5.13}
\frac{1}{2}(-m^2c^4\varphi - c^2\hbar^2\square(\varphi)) \circ v
+
v\circ \frac{1}{2}(-m^2c^4\varphi - c^2\hbar^2\square(\varphi))
\\
+ a
( v\circ\varphi^{k-1} + \varphi\circ v\circ\varphi^{k-2} + \dots
	+ \varphi^{k-2}\circ v\circ\varphi + \varphi^{k-1}\circ v )
= 0 .
	\end{multline}
Choosing $v=\id_\Hil$, we obtain a `classical' Euler-Lagrange equation for
$\varphi$:
	\begin{equation}	\label{5.14}
m^2c^4\varphi + c^2\hbar^2\square(\varphi) = a k \varphi^{k-1} .
	\end{equation}
Combining~\eref{5.14} with~\eref{5.13}, we see that the variation
$v=\delta_0\varphi$ \emph{cannot be arbitrary} (for $a\not=0$) as it must
satisfy the equation
	\begin{equation}	\label{5.15}
\bigl( 1- {k}/{2} \bigr) ( \varphi^{k-1}\circ v + v\circ\varphi^{k-1} )
+
\varphi\circ v\circ\varphi^{k-2} + \varphi^2\circ v\circ\varphi^{k-3}
+ \dots +
\varphi^{k-2}\circ v \circ\varphi
= 0
	\end{equation}
which always has solutions of the form
$v=f(x)\id_\Hil$ with $f\colon M\to\field[C]$. In particular, for $k=1,2,3,4$
this equation respectively reads:
	\begin{equation}	\label{5.16}
	\begin{split}
& 0 = 0 \quad(k=1) \qquad 0 =0 \quad (k=2)
\\
& [\varphi,[\varphi,v]_{\_}]_{\_} = 0 \quad(k=3)
\qquad
[\varphi^2,[\varphi,v]_{\_}]_{\_} = 0 \quad(k=4) .
	\end{split}
	\end{equation}
Consequently, for $k\ge3$ either the set of the variation $v$ should be
restricted to the ones satisfying~\eref{5.15} with $\varphi$ being a
solution~\eref{5.14}, or to the field equation~\eref{5.14} should be added
the condition~\eref{5.15} with arbitrary operator/variation $v$.

\subsection{Interacting neutral scalar fields}
\label{Subsect5.4}

	Consider a system of two neutral scalar fields $\varphi_1$ and
$\varphi_2$ with Lagrangian
	\begin{multline}	\label{5.17}
L
=
- \frac{1}{2}m^2c^4\varphi_1\circ\varphi_1
+ \frac{1}{2}c^2\hbar^2(\pd_\mu\varphi_1)\circ (\pd^\mu\varphi_1)
\\- \frac{1}{2}m^2c^4\varphi_2\circ\varphi_2
+ \frac{1}{2}c^2\hbar^2(\pd_\mu\varphi_2)\circ (\pd^\mu\varphi_2)
+ a\varphi_1\varphi_2
	\end{multline}
where $a$ is a non-vanishing real parameter. Performing a procedure similar
to the ones in the previous subsections, we obtain the field equations
	\begin{gather}	\label{5.18}
m^2c^4\varphi_1 + c^2\hbar^2\square(\varphi_1) = a\varphi_2
\qquad
m^2c^4\varphi_2 + c^2\hbar^2\square(\varphi_2) = a\varphi_1
\\\intertext{and the conditions}
			\label{5.19}
[\varphi_1,v_2]_{\_} = 0
\qquad
[\varphi_2,v_1]_{\_} = 0
	\end{gather}
which must satisfy the solutions of~\eref{5.18} and the variations $v_1$ and
$v_2$ of $\varphi_1$ and $\varphi_2$, respectively. The last conditions are
quite natural from physical view\ndash point because they mean that we can
make a variation of $\varphi_1$ independently of $\varphi_2$ and \emph{vice
versa}. The energy\ndash momentum operator of the system under consideration
turns to be
	\begin{equation}	\label{5.19new}
T_{\mu\nu}
= \frac{1}{2} c^2\hbar^2 \sum_{i=1,2}\bigl\{
(\pd_\mu\varphi_i)\circ(\pd_\nu\varphi_i)
+
(\pd_\nu\varphi_i)\circ(\pd_\mu\varphi_i)
\bigr\}
- \eta_{\mu\nu} L.
	\end{equation}

\subsection{Free Dirac (spinor) field 1. Standard Lagrangian}
\label{Subsect5.5}

	As a standard Lagrangian of a spin $\frac{1}{2}$ Dirac field $\psi$,
we take~\cite{Bogolyubov&Shirkov}
	\begin{equation}	\label{5.19-1}
L
=
\frac{1}{2}\ih c \{
\overline\psi\gamma^\mu \odot (\pd_\mu\psi)
-
(\pd_\mu\overline\psi)\gamma^\mu \odot \psi
\}
- mc^2 \overline\psi \odot \psi.
	\end{equation}
Here: $\gamma^\mu$ are the Dirac's $\gamma$-matrices,
$\overline\psi:=\psi^\dag\gamma^0$ is the Dirac conjugate of
$\psi=(\psi_0,\psi_1,\psi_2,\psi_3)^\top$ with $\top$ being the matrix
transposition sign, in products like $\overline\psi\gamma^\mu$ the matrix
multiplication sign is dropped, and $\odot$ denotes a composition combined with
matrix multiplication, e.g.
\(
\overline\psi \odot \psi
= \sum_\mu \overline\psi_\mu \circ \psi_\mu
= (\gamma^0)^{\mu\nu} \psi_\mu^\dag \circ \psi_\nu .
\)
If $v$ and $\overline v$ denote variations of $\psi$ and $\overline\psi$,
respectively, we, in view of definition~\ref{Defn4.1}, obtain
	\begin{gather}
			\label{5.19-2}
	\begin{split}
\frac{\pd L}{\pd\psi} (v)
& =
- \frac{1}{2}\ih c (\pd_\mu\overline\psi)\gamma^\mu\odot v
	- mc^2 \overline\psi\odot v
\qquad
\frac{\pd L}{\pd(\pd_\mu\psi)} (v)
=
  \frac{1}{2}\ih c \overline\psi \gamma^\mu\odot v
\\
\frac{\pd L}{\pd\overline\psi} (\overline v)
&=
  \frac{1}{2}\ih c \overline v\odot \gamma^\mu (\pd_\mu\psi)
	- mc^2\overline v\odot \psi
\qquad
\frac{\pd L}{\pd(\pd_\mu\overline\psi)} (\overline v)
=
- \frac{1}{2}\ih c \overline v\odot \gamma^\mu\psi .
	\end{split}
\intertext{Thus, the basic relation~\eref{4.13} takes the form}
			\label{5.19-3}
- \{ \ih c (\pd_\mu\overline\psi)\gamma^\mu + mc^2 \overline\psi \} \odot v
+
\overline v\odot \{ \ih c \gamma^\mu (\pd_\mu\psi) - mc^2 \psi \}
= 0 .
	 \end{gather}
The choices when $v$ or $\overline v$ is the zero operator and
$\overline v$ or $v$, respectively, is arbitrary, result in the Dirac
equation and its conjugate,%
\footnote{~%
More precisely, the choice when all but one of the components $v_\alpha$ and
$\overline{v}_\alpha$ vanish, results in the (conjugate) Dirac equation for
this component; the substitution of these results into~\eref{5.19-3} converts
it into identity relative to $v_\alpha$ and $\overline{v}_\alpha$.%
}
\viz
	\begin{equation}
			\label{5.19-4}
\ih  \gamma^\mu (\pd_\mu\psi) - mc \psi = 0
\qquad
\ih  (\pd_\mu\overline\psi)\gamma^\mu + mc \overline\psi= 0 .
	\end{equation}
These equations convert~\eref{5.19-3} into identity relative to $v$ and
$\overline v$.  So, the variations of a free spinor field are completely
arbitrary.  Combining~\eref{5.19-2} with~\eref{4.17}--\eref{4.20}, we get
the energy\ndash momentum, current and spin angular momentum operators as:
	\begin{align}
			\label{5.19-5}
T_{\mu\nu}
& =
\frac{1}{2} \ih c \{
\overline\psi\gamma_\mu\odot (\pd_\nu\psi)
- (\pd_\nu\overline\psi)\gamma_\mu\odot \psi \}
\\			\label{5.19-6}
J_\mu
& = q c \overline\psi\gamma_\mu \odot \psi
\\			\label{5.19-7}
S_{\mu\nu}^{\lambda}
& =
- \frac{1}{4}\hbar c
\overline\psi
( \gamma^\lambda \sigma_{\mu\nu} + \sigma_{\mu\nu} \gamma^\lambda )
\odot \psi
\qquad
\sigma_{\mu\nu}
:= \frac{\iu}{2} (\gamma^\mu\gamma^\nu - \gamma^\nu\gamma^\mu) ,
	\end{align}
where we have used that a 4-rotation
$x^\mu\mapsto x^\mu + \varepsilon^{\mu\nu} x_\nu$ implies
 $\psi\mapsto \e^{-\frac{\iu}{4}\sigma_{\mu\nu}\varepsilon^{\mu\nu}} \psi$
and
\(
\overline\psi\mapsto
\e^{+\frac{\iu}{4}\sigma_{\mu\nu}\varepsilon^{\mu\nu}} \overline\psi .
\)
These expressions are identical with the ones in~\cite{Bogolyubov&Shirkov}.

\subsection
[Free Dirac (spinor) field 2. Charge symmetric Lagrangian]
{Free Dirac (spinor) field\\ 2. Charge symmetric Lagrangian}
\label{Subsect5.5new}

	In the present subsection, we shall present an example of a
Lagrangian for which the Euler\ndash Lagrange equations are identities, like
$0=0$, but which, regardless of this `strange' fact, entails completely
reasonable field equations. In a classical sense, this Lagrangian is a
`completely singular' one, as all its classical derivatives are zero, but it
is not a constant operator.

	Let $\psi$ be a Dirac 4-spinor and (see,
e.g.,~\cite{Bogolyubov&Shirkov,Bjorken&Drell,Roman-QFT})
	\begin{gather}	\label{12.12}
\bpsi := C\opsi^\top = (\opsi C^\top)^\top
\intertext{be its \emph{charge conjugate} one, where the matrix $C$ satisfies
the conditions}
			\label{12.13}
C^{-1} \gamma^\mu C = -\gamma^{\mu\,\top}:=-(\gamma^\mu)^\top
\qquad
C^\top = - C.
	\end{gather}
Let us consider $\psi$ and $\bpsi$ as independent field variables. In their
terms, the Lagrangian~\eref{5.19-1} reads
	\begin{equation}	\label{12.14}
\tope{L}^{\prime}
=
- \frac{1}{2}\ih c\{
  \tope{\bpsi}^\top(x) C^{-1}\gamma^\mu\odot(\pd_\mu\tope{\psi}(x))
- (\pd_\mu\tope{\bpsi}^\top(x)) C^{-1}\gamma^\mu\odot \tope{\psi}(x)
\}
+ mc^2 \tope{\bpsi}^\top(x) C^{-1}\odot \tope{\psi}(x) .
	\end{equation}
We would like to emphasize on the change of the signs and the appearance
of the matrix $C$ in~\eref{12.14} with respect to~\eref{5.19-1}. An
alternative to this Lagrangian is a one with changed positions of $\psi$ and
$\bpsi$, \viz
	\begin{equation}	\label{12.15}
\tope{L}^{\prime\prime}
=
- \frac{1}{2}\ih c\{
  \tope{\psi}^\top(x) C^{-1}\gamma^\mu\odot(\pd_\mu\tope{\bpsi}(x))
- (\pd_\mu\tope{\psi}^\top(x)) C^{-1}\gamma^\mu\odot \tope{\bpsi}(x)
\}
+ mc^2 \tope{\psi}^\top(x) C^{-1}\odot \tope{\bpsi}(x) .
	\end{equation}
	Evidently, the variables $\psi$ and $\bpsi$ do not enter
in~\eref{12.14} and~\eref{12.15} on equal footing.  We shall try to
`symmetrize' the situation by considering a Lagrangian which is the half sum
of the last two ones, \ie
	\begin{multline}	\label{12.16}
\tope{L}^{\prime\prime\prime}
=
  \frac{1}{4}\ih c\{
- \tope{\bpsi}^\top(x) C^{-1}\gamma^\mu\odot(\pd_\mu\tope{\psi}(x))
+ (\pd_\mu\tope{\bpsi}^\top(x)) C^{-1}\gamma^\mu\odot \tope{\psi}(x)
\\
-  \tope{\psi}^\top(x) C^{-1}\gamma^\mu\odot(\pd_\mu\tope{\bpsi}(x))
+ (\pd_\mu\tope{\psi}^\top(x)) C^{-1}\gamma^\mu\odot \tope{\bpsi}(x)
\}
\\
+ \frac{1}{2} mc^2
\{	\tope{\bpsi}^\top(x) C^{-1}\odot \tope{\psi}(x)
      +	\tope{\psi}^\top(x) C^{-1}\odot \tope{\bpsi}(x)
\} .
	\end{multline}

	Let $v$ and $\bv$ denote variations of $\psi$ and $\bpsi$,
respectively. Applying definition~\ref{Defn4.1}, we can calculate the
derivatives of the Lagrangians~\eref{12.14}--\eref{12.16}. They are as
follows:
	\begin{gather*}	\label{5.19-8}
	\begin{split}
\frac{\pd L^{\prime}}{\pd\psi^\top} (v)
& =
+ \frac{1}{2}\ih c \{ C^{-1}\gamma^\mu (\pd_\mu\bpsi) \}^\top \odot v
	- mc^2 \{ C^{-1} \bpsi \}^\top \odot v
\\
\frac{\pd L^{\prime}}{\pd\bpsi^\top} (\bv)
&=
- \frac{1}{2}\ih c \bv^\top \odot C^{-1}\gamma^\mu (\pd_\mu\psi)
	+ mc^2 v^\top \odot C^{-1} \psi
\\
\frac{\pd L^{\prime}}{\pd(\pd_\mu\psi^\top)} (v)
& =
- \frac{1}{2}\ih c \{ C^{-1}\gamma^\mu  \bpsi \}^\top  \odot v
\qquad \!\!\!\!\!
\frac{\pd L^{\prime}}{\pd(\pd_\mu\bpsi^\top)} (\bv)
=
+ \frac{1}{2}\ih c \bv^\top \odot C^{-1}\gamma^\mu\psi
	\end{split}
\displaybreak[1]\\
	\begin{split}
\frac{\pd L^{\prime\prime}}{\pd\psi^\top} (v)
& =
- \frac{1}{2}\ih c v^\top \odot C^{-1}\gamma^\mu (\pd_\mu\bpsi)
	+ mc^2  v^\top \odot C^{-1} \bpsi
\\
\frac{\pd L^{\prime\prime}}{\pd\bpsi^\top} (\bv)
&=
+ \frac{1}{2}\ih c \{ C^{-1}\gamma^\mu (\pd_\mu\psi) \}^\top \odot \bv
	- mc^2 \{ C^{-1} \psi \}^\top \odot \bv
\\
\frac{\pd L^{\prime\prime}}{\pd(\pd_\mu\psi^\top)} (v)
& =
+ \frac{1}{2}\ih c v^\top \odot C^{-1}\gamma^\mu  \bpsi
\qquad \!\!\!\!\!
\frac{\pd L^{\prime\prime}}{\pd(\pd_\mu\bpsi^\top)} (\bv)
=
- \frac{1}{2}\ih c \{ C^{-1}\gamma^\mu\psi \}^\top \odot \bv
	\end{split}
\displaybreak[1]\\
	\begin{split}
2 \frac{\pd L^{\prime\prime\prime}}{\pd\psi^\top} (v)
& =
+ \frac{1}{2}\ih c \{ C^{-1}\gamma^\mu (\pd_\mu\bpsi) \}^\top \odot v
	- mc^2 \{ C^{-1} \bpsi \}^\top \odot v
\\ & \hphantom{=}
- \frac{1}{2}\ih c v^\top \odot C^{-1}\gamma^\mu (\pd_\mu\bpsi)
	+ mc^2  v^\top \odot C^{-1} \bpsi
\\
2 \frac{\pd L^{\prime\prime\prime}}{\pd\bpsi^\top} (\bv)
&=
- \frac{1}{2}\ih c \bv^\top \odot C^{-1}\gamma^\mu (\pd_\mu\psi)
	+ mc^2 v^\top \odot C^{-1} \psi
\\ & \hphantom{=}
+ \frac{1}{2}\ih c \{ C^{-1}\gamma^\mu (\pd_\mu\psi) \}^\top \odot \bv
	- mc^2 \{ C^{-1} \psi \}^\top \odot \bv
\\
2 \frac{\pd L^{\prime\prime\prime}}{\pd(\pd_\mu\psi^\top)} (v)
& =
- \frac{1}{2}\ih c \{ C^{-1}\gamma^\mu  \bpsi \}^\top  \odot v
+ \frac{1}{2}\ih c v^\top \odot C^{-1}\gamma^\mu  \bpsi
\\
2 \frac{\pd L^{\prime\prime\prime}}{\pd(\pd_\mu\bpsi^\top)} (\bv)
& =
+ \frac{1}{2}\ih c \bv^\top \odot C^{-1}\gamma^\mu\psi
- \frac{1}{2}\ih c \{ C^{-1}\gamma^\mu\psi \}^\top \odot \bv .
	\end{split}
	\end{gather*}
Notice, in these equalities the matrix transposition serves only to ensure
proper matrix multiplication.
	Substituting the just calculated derivatives in the basic
equation~\eref{4.13}, we see that for the
Lagrangians~\eref{12.14}--\eref{12.16} it reduces respectively to
	\begin{align}	\label{5.19-30}
+ \bigl( A(\bpsi) \bigr)^\top \odot v
- \bv^\top \odot A(\psi)
& = 0
\\			\label{5.19-31}
- v^\top \odot A(\bpsi)
+ \bigl( A(\psi) \bigr)^\top \odot \bv
& = 0
\\			\label{5.19-32}
+ \bigl( A(\bpsi) \bigr)^\top \odot v
- v^\top \odot A(\bpsi)
+ \bigl( A(\psi) \bigr)^\top \odot \bv
- \bv^\top \odot A(\psi)
& = 0
	\end{align}
where
		\begin{equation}	\label{5.19-33}
A\colon \psi \mapsto
A(\psi)
:=
\ih c C^{-1}\gamma^\mu \pd_\mu \psi - m c^2 C^{-1}\psi
	\end{equation}
Let $\alpha\in\{0,1,2,3\}$ be arbitrarily fixed.
Choosing all but the $\alpha^{\text{th}}$ component of $v=0$ (resp.\ $\bv=0$)
to vanish and setting this component to be equal to the identity mapping
$\id_\Hil$, we see that~\eref{5.19-30} or~\eref{5.19-31} is equivalent to
the equations $A(\psi_\alpha)=0$ and $A(\bpsi_\alpha)=0$ for all
$\alpha=0,1,2,3$, which, by virtue of~\eref{5.19-33}, are equivalent to the
system of Dirac equations
		\begin{equation}	\label{5.19-34}
\ih \gamma^\mu \pd_\mu \psi - m c^2 \psi = 0
\qquad
\ih \gamma^\mu \pd_\mu \bpsi - m c^2 \bpsi = 0 .
	\end{equation}
However, if one substitutes in~\eref{5.19-32} the just-described choices of
$v$ and $\bv$, one will get the identity $0=0$ for any one of them, instead
of some equations for the components of $\psi$ and $\bpsi$. In this sense,
\emph{classical Euler\ndash Lagrange equations for the
Lagrangian~\eref{12.16} do not exist}.  In this case, which cannot be handled
by the standard methods, we shall proceed as follows. Let us make the same as
the above selection of the variations $v$ and $\bv$ without setting the
non\ndash vanishing components to be equal to $\id_\Hil$. Such choices
reduces~\eref{5.19-32} to the next system of relations (do not sum over
$\alpha$!)
	\begin{equation}	\label{5.19-35}
[A(\bpsi_\alpha),v_\alpha]_{\_} = 0
\quad
[A(\psi_\alpha),\bv_\alpha]_{\_} = 0
	\end{equation}
for the variations $v_\alpha$ and $\bv_\alpha$. If we now choose the
operators $v_\alpha$ and $\bv_\alpha$ to range in a unitary representation in
$\Hil$ of some group, then the Schur's lemma%
\footnote{ %
See, e.g,~\cite[appendix~II]{Rumer&Fet}, or \cite[sec.~8.2]{Kirillov-1976},
or~\cite[ch.~5, sec.~3]{Barut&Roczka}.%
}
implies the existence of classical functions, not operators,
$f_{\psi,\alpha}(x)$ and $f_{\bpsi,\alpha}(x)$, such that
	\begin{equation}	\label{5.19-36}
 A(\psi_\alpha(x)) =  f_{\psi,\alpha}(x) \id_\Hil
\quad
A(\bpsi_\alpha(x)) = f_{\bpsi,\alpha}(x) \id_\Hil.
	\end{equation}
In view of~\eref{5.19-33}, these equations are equivalent to the system
		\begin{equation}	\label{5.19-37}
\ih \gamma^\mu \pd_\mu \psi - m c^2 \psi = \chi_\psi \id_\Hil
\qquad
\ih \gamma^\mu \pd_\mu \bpsi - m c^2 \bpsi = \Breve{\chi}_{\bpsi} \id_\Hil .
	\end{equation}
where $\chi_\psi$ and $\Breve{\chi}_{\bpsi}$ are some \emph{classical}, not
operator\ndash valued, spinors. It is trivial to be checked
that~\eref{5.19-37} converts~\eref{5.19-32} into identity with respect to
$v$ and $\bv$. Therefore the system of (Dirac equations with, generally,
non\ndash vanishing r.h.s.)~\eref{5.19-37} plays a role of a system of
field equations for the Lagrangian~\eref{12.16}.

	Ending this example, we note that if one wants the
Lagrangian~\eref{12.16} to describe a free spinor field, the choices
	\begin{equation}	\label{5.19-38}
\chi_\psi = 0 \quad \Breve{\chi}_{\bpsi} = 0
	\end{equation}
should be made; otherwise, the equations~\eref{5.19-37} describe a spin
$\frac{1}{2}$ quantum field with some selfinteraction. Elsewhere we shall
demonstrate that the Lagrangians~\eref{12.14}--\eref{12.16} and the
additional conditions~\eref{5.19-38} lead to one and the same quantum field
theory of free spinor fields.

\subsection{General quadratic Lagrangian}
\label{Subsect5.6}

	Let us consider a system of quantum fields $\varphi_i$ with a
Lagrangian
	\begin{equation}	\label{5.20}
L
=
  a^i \varphi_i
+ m^{ij} \varphi_i \circ \varphi_j
+ b^{i\mu} \pd_\mu\varphi_i
+ g^{i\mu j\nu} (\pd_\mu\varphi_i) \circ (\pd_\nu\varphi_j)
+ c^{ij \mu} \varphi_i \circ (\pd_\mu\varphi_j)
+ d^{ij \mu} (\pd_\mu\varphi_j) \circ \varphi_i,
	\end{equation}
where $a^i$,  $m^{ij}$, $b^{i\mu}$, $g^{i\mu j\nu}$, $c^{ij \mu}$, and
 $d^{ij \mu}$ are some (dimensional) constants. Calculating
$\frac{\pd L}{\pd\varphi_i}$ and $\pi^{i\mu}=\frac{\pd
L}{\pd(\pd_\mu\varphi_i)}$, according to definition~\ref{Defn4.1}, and
substituting them into~\eref{4.13}, we get the relation
	\begin{gather}
			\label{5.21}
\alpha^i\circ v_i + v_i\circ \beta^i + a^i v_i = 0
\intertext{where}
			\label{5.22}
	\begin{split}
\alpha^i
& :=
m^{ij} \varphi_j + (c^{ij\mu} - d^{ji\mu}) \pd_\mu\varphi_j
	- g^{i\mu j\nu} \pd_\mu\pd_\nu\varphi_j
\\
\beta^i
& :=
m^{ji} \varphi_j + (d^{ij\mu} - c^{ji\mu}) \pd_\mu\varphi_j
	- g^{j\nu i\mu} \pd_\mu\pd_\nu\varphi_j
	\end{split}
	\end{gather}
and $v_i=\delta_0\varphi_i$ is a variation of $\varphi_i$. Choosing
$v_i=f_i(x)\id_\Hil$ with arbitrary $f\colon M\to\field[C]$,
from~\eref{5.21}, we obtain the Euler\ndash Lagrange equations
	\begin{gather}
			\label{5.23}
\alpha^i +\beta^i + a^i\id_\Hil = 0 .
\intertext{The combination of~\eref{5.23} and~\eref{5.21} results in the
condition}
			\label{5.24}
\sum_{i} [\alpha^i, v_i]_{\_} = 0
\intertext{(which is equivalent to $\sum_{i} [\beta^i, v_i]_{\_} = 0$) for
the field operators $\varphi_i$ and their variations $v_i$. These conditions
can be satisfied identically relative to $v_i$ if all $\alpha^i$ (and, hence,
all $\beta^i$) happen to be proportional to the identity operator $\id_\Hil$.
In particular, this will be the case when}
			\label{5.25}
m^{ij} = m^{ji} \quad
c^{ij\mu} - d^{ji\mu} = d^{ij\mu} - c^{ji\mu} \quad
g^{i\mu j\nu} = g^{j\nu i\mu},
\intertext{so that $\alpha^i=\beta^i$ (see~\eref{5.22}) and the field
equations~\eref{5.23} read}
			\label{5.26}
\alpha^i + \frac{1}{2} a^i\id_\Hil = 0 .
\\\intertext{As a special case of~\eref{5.25}, the one of free fields should
be single out. In this important case, we have (do not sum over $i$!)}
			\label{5.27}
a^i = b^i = c^{ij\mu} = d^{ij\mu} = 0 \quad
m^{ij} = m_i^2 c^4 \delta^{ij} \quad
g^{i\mu j\nu} = g^{\mu\nu} \delta^{ij}
	\end{gather}
where $m_i$ and $g^{\mu\nu}=g^{\nu\mu}$ are some (dimensional)
constants.

	However, having in mind the considerations in
Subsect.~\ref{Subsect5.5new}, one should take into account the possibility
that the Euler\ndash Lagrange equations~\eref{5.23} may turn to be
identities, in which case the field equations should be derived
from~\eref{5.21} in a different way.

	The energy-momentum operator corresponding to the
Lagrangian~\eref{5.20}, in view of~\eref{4.17}, is
	\begin{multline}	\label{5.28}
T^{\mu\nu}
=
b^{i\mu} (\pd^\nu\varphi_i)
+ g^{i\mu j\lambda} (\pd^\nu\varphi_i) \circ (\pd_\lambda\varphi_j)
+ g^{j\lambda i\mu} (\pd_\lambda\varphi_j) \circ (\pd^\nu\varphi_i)
\\
+ c^{ji \mu} \varphi_j \circ (\pd^\nu\varphi_i)
+ d^{ji\mu} (\pd^\nu\varphi_i) \circ \varphi_j
- \eta^{\mu\nu} L .
	\end{multline}
If the system possesses a charge, the corresponding current operator can be
obtained formally from the r.h.s.\ of~\eref{5.28} by deleting the last term
and replacing $\pd^\nu$ with $\frac{q}{\ih c}\varepsilon(\varphi_i)$
(see~\eref{4.18}).


\section {Conclusion}
\label{Conclusion}

	In this paper we have given an analysis of	 some aspects and
corollaries of the Schwinger's action principle in (canonical) quantum field
theory. As it was demonstrated, in the `standard' presentation, this
variational principle contains an additional hypothesis, which does not
follow logically from the rest of the theory and modifies the conserved
quantities so that they do not always have the required properties. We have
removed the mentioned hypothesis by giving a suitable meaning of derivatives
of operator\ndash valued functions of operator arguments with respect to such
an argument.  After that modification is done, the following important
consequences of the Schwinger's action principle were derived:

\renewcommand{\theenumi}{\roman{enumi}}
	\begin{enumerate}
\item
The classical (standard) Euler-Lagrange equation of motion for the field
operators remain the same as before the change.

\item
The conserved quantities (operators) are changed so that they have the
required properties, at least in the examples considered.

\item	\label{VariationEquation}
In the general case (of (self-)interacting fields), the variations of the
field operators cannot be completely arbitrary as they should satisfy some
equations in which the field operators, satisfying the Euler\ndash Lagrange
equations, enter.

\item	\label{IdVariations}
Any variations of the field operators proportional to the identity
mapping/operator, like~\eref{3.25}, are always admissible.

\item
The variations if item~\ref{IdVariations} above are sufficient for the
derivation of the Euler-Lagrange equations and all spacetime conserved
quantities. However, for other variations, such as the ones connected with
internal symmetries (e.g.\ like (constant) phase transformations leading to
charge conservation), one should always check whether they are admissible in a
sense that they must satisfy the equations mentions in
point~\ref{VariationEquation} above.

\item
If one insists on keeping the field variations completely arbitrary, it is
quite likely that, for interacting fields, field operators which are multiples
of the identity operator will be the only solutions of the variational
problem determining them.%
\footnote{~%
For instance, such is the case considered in Subsect.~\ref{Subsect5.4}:
if~\eref{5.19} holds for any $v_1$ and $v_2$, then
$\varphi_a(x)=f_a(x)\id_\Hil$, $a=1,2$, for some $f_a\colon M\to\field[C]$
(as a result of, e.g., Schur's lemma ---see~\cite[sec.~8.2]{Kirillov-1976} or
\cite[ch.~5, sec.~3]{Barut&Roczka}).%
}
	\end{enumerate}

	At the end, since the Euler-Lagrange equations are not changed after
the described correction of Schwinger's action principle, the modification in
the structure of conserved quantities (operators), and, possibly, other
equations for the fields (and their variations) should be regarded as the
main outcome of the present investigation.




\addcontentsline{toc}{section}{References}
\bibliography{bozhopub,bozhoref}
\bibliographystyle{unsrt}
\addcontentsline{toc}{subsubsection}{This article ends at page}

\end{document}